\documentclass{aa}  
\usepackage{ulem}
\usepackage{graphicx}
\usepackage{txfonts}
\begin{document} 

   \title{Trapping planets in an evolving protoplanetary disk: preferred time, locations, and planet mass.}


   \author{K. Bailli\'e
          \inst{1,2,3}
          \and
          S. Charnoz\inst{2,3}
          \and
          E. Pantin\inst{3}
          }

   \institute{IMCCE, Observatoire de Paris, PSL Research University, CNRS, Sorbonne Universit\'es, UPMC, Univ. Lille 1, 77 Av. Denfert-Rochereau, 75014 Paris, France.\\
         \and
			Institut de Physique du Globe, Sorbonne Paris Cit\'e, Universit\'e Paris Diderot/CNRS, 1 rue Jussieu, 75005 Paris, France.\\
         \and
			Laboratoire AIM-LADP, Universit\'e Paris Diderot/CEA/CNRS, 91191 Gif sur Yvette, France.\\
\\
              \email{kevin.baillie@obspm.fr}
             }

   \date{Accepted March 2016}

 
  \abstract
{Planet traps are necessary to prevent forming planets from falling onto their host star by type I inward migration. Surface mass density and temperature gradient irregularities favor the apparition of traps (planet accumulation region) and deserts (planet depletion zone). These features are found at the dust sublimation lines and heat transition barriers.}
   {We study how planets may remain trapped or escape these traps as they grow and as the disk evolves viscously with time.}
   {We numerically model the temporal viscous evolution of a protoplanetary disk by coupling its dynamics, thermodynamics, geometry, and composition. The resulting mid-plane density and temperature profiles allow the modeling of the interactions of this type of evolving disk with potential planets, even before the steady state is reached.}
   {We follow the viscous evolution of a minimum mass solar nebula (MMSN) and compute the Lindblad and corotation torques that this type of disk would exert on potential planets of various masses that are located within the planetary formation region. We determine the position of planet traps and deserts in relationship with the sublimation lines, shadowed regions, and heat transition barriers. We notice that the planet mass affects the trapping potential of the mentioned structures through the saturation of the corotation torque. Planets that are a few tens of Earth masses can be trapped at the sublimation lines until they reach a certain mass while planets that are more massive than $100 M_{\mathrm{\oplus}}$ can only be trapped permanently at the heat transition barriers. They may also open gaps beyond 5 au and enter type II migration.}
   {Coupling a bimodal planetary migration model with a self-consistent evolved disk, we were able to distinguish several potential planet populations after five million years of evolution: two populations of giant planets that could stay trapped around 5.5 and 9 au and possibly open gaps, some super-Earths trapped around 5 and 7.5 au, and a population of close-in super-Earths, which are trapped inside 1 au. The traps that correspond to the last group could help to validate the in situ formation scenarios of the observed close-in super-Earths.}


\defcitealias{baillie14}{BC14}
\defcitealias{baillie15}{BCP15}

   \maketitle
%

\section{Introduction} \label{intro}

With the newest telescopes (either on the ground or in space), we now have access to a huge variety of observations of both protoplanetary disks and exoplanets. These observations combine to give a more global vision of planetary formation. It is very important, therefore, to understand how the disks select the planets that are to survive, but also how the planets reshape the disks. To achieve that, it is necessary to understand the disks' structure, their evolution, and their interactions with the planets they host.

Observations by \citet{beckwith96} and \citet{hartmann98} estimated a typical disk lifetime of a few million years, consistent with the time required for gaseous planet formation involving gas accretion on solid cores \citep{pollack96}. However, \citet{papaloizou99} suggested that hot Jupiters may have formed at different locations than the ones where they are observed, involving some planetary migration. Long-known type I Lindblad inward migration was extensively described in \citet{goldreich79, arty93} and \citet{ward97}, while \citet{ward91} added an analytical description of the corotation torque. In particular, they estimated a migration timescale of a typical Earth-mass planet in a minimum mass solar nebula (MMSN) \citep{weiden77, hayashi81} around $10^{5}$ years. \citet{kory93}, \citet{alibert05} and \citet{ida08} note an inconsistency between the planet formation and migration timescales: planets should not have time to grow before they fall spiraling into their host star. Various attempts at slowing down the cores' inward migration invoked a regular magnetic field \citep{terquem03}, turbulent magnetism \citep{nelson04}, inner cavities \citep{kuchner02}, self-shadowing effects \citep{jc05} or strong positive density gradients \citep{masset06b}. \citet{paarm08,paarp08,baruteau08} derive analytical expressions from numerical simulations to model inward migration, while accounting for the disk profile and, in particular, its thermodynamical structure. \citet{menou04} estimate that opacity transitions could alter and even stop inward migration in steady state viscous $\alpha$ disks. Planets at these locations would be trapped, preventing their fall onto the star and favoring planet growth by creating a region where trapped planetary embryos would accumulate and collide. \citet{paarp09a} analyze the possibility to trap planets with positive surface mass density gradients, while \citet{lyra10,bitsch11,hasegawa112} estimate the trap locations.

Most of the previous analysis of planet migration was enabled by modeling simplified disk structures using power-law prescriptions for the disk mid-plane density and temperature \citep{hasegawa111,paar11}, density prescriptions with self-consistent 2D-temperature structure \citep{bitsch11,bitsch13} or steady-state accretion disk models \citep{bitsch14}. Though these disk models appear simpler on large scales,  these approaches permitted the development of locally valid torque expressions, which are now widely used for modeling planetary migration. In this paper, we use the disk structure generated by the hydrodynamical numerical simulations described in \citet{baillie14} and \citet{baillie15} (referred hereafter as \citetalias{baillie14} and \citetalias{baillie15}). Coupling the disk dynamics, its geometry, and its thermal structure, this code models the long-term viscous evolution of $\alpha$ disks \citep{shakura73} and is able to numerically retrieve some observational characteristics of protoplanetary disks. Based on \citet{helling00} and \citet{,semenov03}'s opacity tables for the main elements that make up the disk dust, \citetalias{baillie15}'s code relates the disk temperature to the disk dust composition and dust-to-gas ratio at that temperature. From the obtained disk structures in density and temperature, it is possible to derive the torques exerted by the disk on putative planets, that induce their inward or outward migration. These torques are particularly sensitive to the gradients of density and temperature. This leads logically to migration changes at the places where the gradients are most affected: at the sublimation lines of the main components of the dust and at the heat transition barrier that separates the inner disk under viscous heating domination and the outer disk that is dominated by stellar irradiation heating (as suggested by \citet{hasegawa112}). \citetalias{baillie15} found that $10 M_{\mathrm{\oplus}}$-planets could get trapped at these locations.

In this paper, we investigate how the planet mass affects its ability to get trapped and saved. We also analyze how growing planets may remain trapped as the disk evolves. We show that only intermediate-mass planets (a few tens of Earth masses) can get trapped at the sublimation lines (mainly the water ice, refractory organics and silicates lines) whereas the heat transition barrier might save more massive planets. However, the most massive planets at that heat frontier might open gaps and move from type I to type II migration if they become more massive than $\sim 170 M_{\mathrm{\oplus}}$. In addition, our disk simulations show the possibility for super-Earths to get trapped in several specific locations inside 2 au. The distribution of some of these traps could be related to the distribution of resonant pairs seen in exoplanet observations (\citet{schneider11,wright11}). The super-Earth traps we found may also provide a plausible way to allow the in situ formation scenario of super-Earths inside 1 au, as described by \citet{ogihara15}.

Section \ref{methods} details how \citet{baillie15}'s hydrodynamical code allows the viscous evolution of a protoplanetary disk to be tracked by accounting for the sublimation of the main components of the dust. Section \ref{res} describes the interaction between the disk and potential planets, calculates the torques, and determines the locations of the planet traps and deserts. The influence of the planet mass on its ability to get trapped is detailed in Section \ref{disc}. The impact of planetary growth and the possibility of opening a gap are investigated in Section \ref{trappinggrowingplanets}, while the impact of the inner planetary traps on the close-in super-Earth formation scenarios is investigated in Section \ref{se}. Finally, Section \ref{cclpersp} draws conclusions and details some perspectives to better model how planetary growth will affect the disk.

\section{Methods} \label{methods}
\subsection{Disk evolution model}

Following the model described in \citetalias{baillie14} and \citetalias{baillie15}, we consider a viscous $\alpha$ disk \citep{shakura73} and we use similar terminology to that used in those papers. We set the turbulent viscosity to $\alpha_{\mathrm{visc}}= 10^{-2}$, as is commonly taken for disks around T Tauri stars in the absence of deadzones \citep{fromang06}. Using the mass and angular momentum conservation, Equation \ref{lb74} from \citet{lyndenbellpringle74} describes the time evolution of the surface mass density.

\begin{equation}
\frac{\partial \Sigma(r,t)}{\partial t} = \frac{3}{r} \, \frac{\partial}{\partial r}\left(\sqrt{r} \, \frac{\partial}{\partial r} \left( \nu(r,t) \, \Sigma(r,t) \, \sqrt{r}\right) \right)
\label{lb74}
,\end{equation}
where $\Sigma$ is the surface mass density and $\nu$ is the viscosity.

Similarly to \citetalias{baillie14} and \citetalias{baillie15}, we define annuli of masses that are logarithmically distributed in radii between $R_{*}$ and 1000 au. We then apply Equation \ref{lb74} to the 1D grid in surface mass density. We set the boundary conditions so that the flux at the inner edge cannot be directed outward (the disk cannot gain mass from the star). The inner mass flux is therefore the mass accretion rate of the disk onto the star.

At each time step and for each annulus, the disk mid-plane temperature, $T_{m}(r)$, is calculated by accounting for stellar irradiation heating, radiative cooling, and viscous heating of the disk material. The iterative process detailed in \citetalias{baillie14}, by which the disk radial temperature profile is determined, estimates the vertical structure of the disk, the shape of the photosphere, and the composition of the disk mid-plane at the same time.

The angle at which the star sees the photosphere at a given radial location $r$ is the grazing angle, $\alpha_{gr}(r)$. Equation \ref{alphagr} describes how the photosphere height $H_{ph}(r)$ is related to this angle, which actually controls the amount of energy that the star provides to the disk photosphere.

\begin{equation}
\alpha_{gr}(r) = \arctan\left(\frac{dH_{ph}}{dr}(r)\right) - \arctan\left(\frac{H_{ph}(r)-0.4 R_{*}}{r}\right)
\label{alphagr}
.\end{equation}

Equation 18 from \cite{calvet91} describes the resulting heating in the mid-plane while the viscous heating contribution depends on the surface mass density and the viscosity (i.e. the mid-plane temperature):

\begin{equation}
F_{v}(r) = \frac{1}{2} \Sigma(r) \nu(r) \left( R \frac{\mathrm{d}\Omega}{\mathrm{d}r} \right)^{2} = \frac{9}{4} \Sigma(r) \nu(r) \Omega^{2}(r).
\end{equation}

Therefore, we solve the energy equation by considering the irradiation heating as an implicit function of the mid-plane temperature. Assuming the vertical distribution of material to follow the Gaussian law that results from a vertical isothermal distribution in hydrostatic equilibrium, we can use Equation A9 from \cite{dullemond01} to calculate the ratio $\chi$ of the photosphere height $H_{ph}(r)$ to the pressure scale height $h_{pr}(r)$. This approximation is reasonable below a few pressure scale heights, where most of the disk mass is located. The irradiation term directly depends on the opacity of the medium, which in turn depends on the physical states of the various components of the disk material, and therefore depends on the disk mid-plane temperature as explained in Section \ref{realopa}. Then, for a given initial grazing angle, a temperature guess sets the disk local composition, its opacities, its pressure scale height, and therefore its photosphere height (through $\chi$). We can then derive $\frac{dH_{ph}}{dr}$ and use Equation \ref{alphagr} to aim at a convergence on the grazing angle after a few iterations. At radial locations where the algorithm converges towards a solution involving a positive grazing angle, the disk photosphere is considered irradiated. Conversely, when the converged solution requires a negative grazing angle, we consider the disk local photosphere as shadowed by inner regions and therefore not directly irradiated by the star: we then use a simpler algorithm that cancels the contribution of the stellar irradiation. Thus, the geometrical structure (grazing angle, photosphere, and pressure heights), the temperature, and the disk composition are determined jointly by iterating numerically: the algorithm is thoroughly described in \citetalias{baillie14}.

\subsection{Disc composition}
\label{realopa}
\citetalias{baillie14} used a very simple model of opacities: either all the dust particles were sublimated or none, depending on whether the local temperature was higher than 1500 K (considered as the sublimation temperature of the silicate dust) or not. \citetalias{baillie15} used a refined model of opacities based on more dust species with optical constants measured in lab experiments, to account for the variations of the dust composition as a function of the local temperature. The computation of the Rosseland mean opacities follows the procedure described by \citet{helling00} and \citet{semenov03}. \citetalias{baillie15} considered the dust grains to be composed of olivine silicate, iron, pyroxene, troilite, refractory and volatile organics, and water ice, with the initial abundances described in Table \ref{tempchgt}. Sublimation temperatures are provided by \citet{pollack94}, corresponding to gas densities of about $10^{-10} \, \mathrm{g/cm^{3}}$. At a given temperature, their opacity model considers a mixture of dust grains with rescaled abundances at that temperature to account for the sublimated components. These opacities vary by several orders of magnitude, in particular around the sublimation temperature of the dust's main components.

\begin{figure}[htbp!]
\center
\includegraphics[width=\hsize]{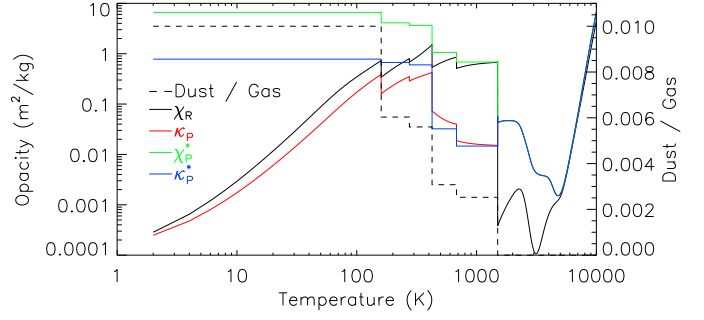}
\caption{Mean-opacity variations with local temperature. Black: Rosseland mean opacity in extinction. Red: Planck mean opacity in absorption. Green: Planck mean opacity in extinction at stellar irradiation temperature. Blue: Planck mean opacity in absorption at stellar irradiation temperature. The corresponding dust-to-gas ratio is displayed with a dashed line.}
\label{opa}
\end{figure}

The Rosseland ($\chi_{R}$) and Planck ($\kappa_{P}$) opacity variations with temperature are presented in Figure \ref{opa} together with the Planck mean opacities at stellar effective temperature $T_{*} = 4000$ K in extinction ($\chi_{P}^{*}$) and absorption ($\kappa_{P}^{*}$).

\begin{table}
\begin{center}
$\begin{array}{c|c|c}
\mathrm{Elements} & \mathrm{Sublimation} & \mathrm{Relative}\\
& \mathrm{temperature} & \mathrm{abundances}\\
\hline
\mathrm{Water\, ice}            &       \mathrm{160\, K}        & 59.46 \, \% \\
\mathrm{Volatile\, organics}    &       \mathrm{275\, K}        & 5.93 \, \% \\
\mathrm{Refractory\, organics}  &       \mathrm{425\, K}        & 23.20 \, \% \\
\mathrm{Troilite\, (FeS)}               &       \mathrm{680\, K}        & 1.57 \, \% \\
\mathrm{Olivine}                &       \mathrm{1500\, K}       & 7.46 \, \% \\
\mathrm{Pyroxene}               &       \mathrm{1500\, K}       & 2.23 \, \% \\
\mathrm{Iron}                   &       \mathrm{1500\, K}       & 0.16 \, \% \\
\end{array}$
\end{center}
\caption{Sublimation temperatures and relative abundances of the disk dust's main components.}
\label{tempchgt}
\end{table}

\section{Results} \label{res}

To better compare our results with previous studies, we decide to model the evolution of a minimum mass solar nebula protoplanetary disk around a classical T Tauri-type young star. Our initial surface mass density profile is given by \citet{weiden77} with the scaling from \citet{hayashi81}:

\begin{equation}
\label{eqmmsn}
\Sigma (r) = 17,000 \left(\frac{r}{1 \, \mathrm{au}}\right)^{-3/2} \mathrm{kg\cdot m^{-2}}
.\end{equation}

The central star is a classical T Tauri-type young star with constant $M_{*} = 1 \, M_{\odot}$, $R_{*} = 3 \, R_{\odot}$, $T_{*} = 4000 \, \mathrm{K}$ and $\mathcal{L}_{*} = 4 \pi \, R_{*}^{2} \, \sigma_{B} \, T_{*}^{4}$ along the simulation.

This choice of initial state is discussed in \citetalias{baillie14} and is also validated by \citet{vorobyov07} who show that the MMSN density profile could be interpreted as an intermediate stage along the evolution of a protoplanetary disk under self-regulated gravitational accretion.

\subsection{Time evolution}
\label{evol}
Though the gas of this type of disk is believed to photo-evaporate by stellar irradiation in a few million years (\citet{font04, alexander07, alexander09, owen10}, \citetalias{baillie15} followed the evolution of this type of disk over 10 million years. 

\begin{figure}[htbp!]
\begin{center} $
\begin{array}{c}
\includegraphics[width=7.5cm, clip=true]{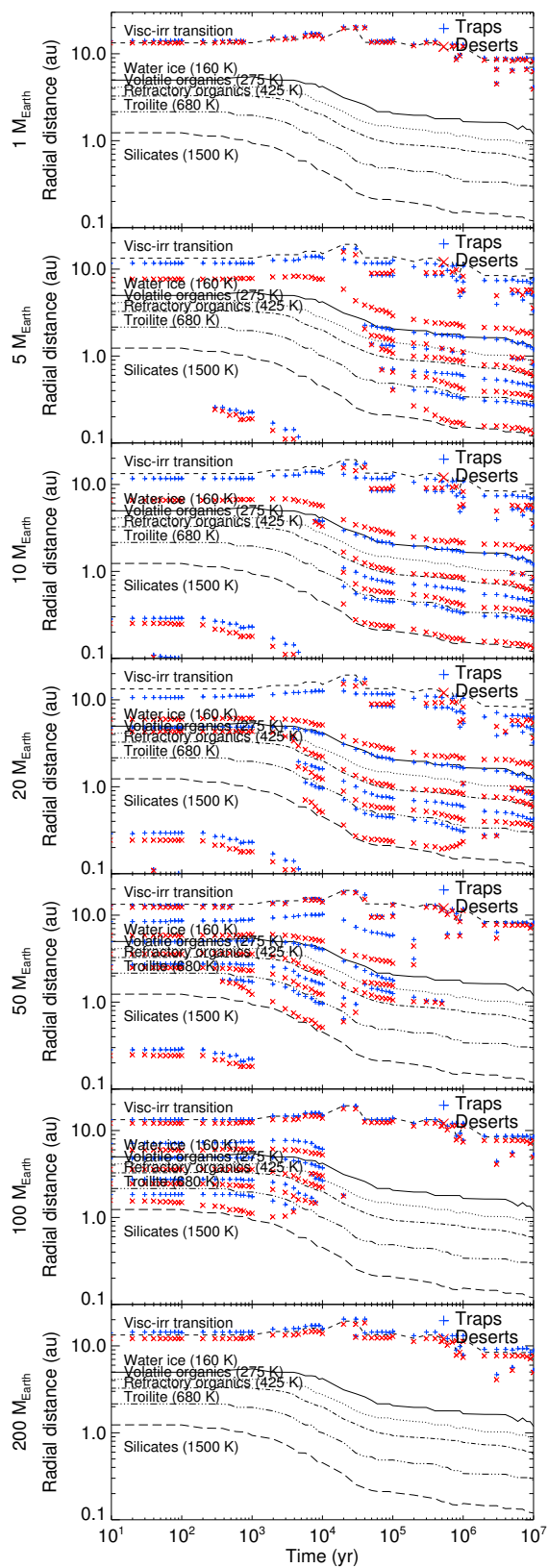}\\
\end{array} $
\end{center}
\caption{Time evolution of the migration traps (blue +) and deserts (red x) positions. The sublimation line positions and the heat transition radius are represented in black. Each subfigure shows the traps and deserts for a given planet mass.}
\label{traptime}
\end{figure}

In addition to following the evolution of the surface mass density, \citetalias{baillie15} also derived the associated mid-plane temperature profiles and noticed temperature plateaux at the sublimation temperatures of the main dust components (referenced in Table \ref{tempchgt}). These plateaux drift inward as the star accretes material form the disk and the disk cools down with time. The mean locations of these sublimation plateaux are referred as "sublimation lines" and are represented with the black dotted and dashed lines in Figure \ref{traptime}.

\subsection{Type I migration}
\label{torque}
The results reported in \citetalias{baillie15} clearly show that radial gradients of surface mass density and temperature that were expected to be mostly negative could strongly vary and even become positive locally when the disk evolves.

Following their methods, we derive the total torques that a viscously evolving disk would give to a putative planet of mass $M_{P}$, which is located at a radial distance $r_{P}$ from the central star. We assume that the planet has already formed. \citet{goldreich79} ,\citet{ward88}, \citet{arty93} ,\citet{jc05} early described that a planet excites resonances in the disk: Lindblad resonances that were caused by the action of the spiral arms were induced by the planet and corotation resonances owing to the horseshoe region around the planet. Since we know the various components of the associated resonant torques (Sections \ref{torquelin}-\ref{torquecor}), assuming that the disk structure is not affected by the planet, we can derive the reaction torque exerted by the disk on the planet.

\citetalias{baillie15} also noticed that putative planets could experience total torques that could change sign when encountering these density and temperature gradient irregularities, which results in planets being possibly trapped at specific locations called "planet traps" and usually located at the outer edge of the sublimation regions of the different dust elements, or at the outer edge of the shadowed regions (regions for which the photosphere is no longer directly irradiated by the star). They also provided evidence that some regions of the disk called "planet deserts" can be totally depleted in planets. Though this was previously discussed in imposed-density or steady-static disks \citep{hasegawa111,paar11,bitsch11,bitsch13,bitsch14}, \citetalias{baillie15} was the first attempt at modeling these traps in viscously evolving disks.

In this paper, we study how these traps and deserts are affected by the mass of the planet. To that purpose, we need to fully account not only for the Lindblad torque but also for the detailed corotation torque, including saturation effects.

\subsubsection{Lindblad and corotation torques}
\label{torquelin}
A planet excites Lindblad resonances of multiple orders in a disk \citep{goldreich79,goldreich80}. Assuming thermal equilibrium and an adiabatic disk, \citet{paarp08} provide a formula for the total Lindblad torque exerted by a 2D laminar disk in the absence of self-gravity on a circular planet. Owing to thermal diffusion, waves propagate at a velocity between the isothermal sound speed (maximum thermal diffusion) and the adiabatic sound speed (no thermal diffusion). \citet{paar11} then generalized the previous formula to account for this difference, therefore replacing the adiabatic index $\gamma$ by an effective index $\gamma_{\mathrm{eff}}$:

\begin{equation}
\label{gammalin}
\gamma_{\mathrm{eff}} \frac{\Gamma_{\mathrm{Lindblad}}}{\Gamma_{0}(r_{P})} = - \left(2.5 \, - 1.7 \frac{\partial \ln T}{\partial \ln r} + 0.1 \frac{\partial \ln \Sigma}{\partial \ln r} \right)_{r_{P}},
\end{equation}
with $q = \frac{M_{planet}}{M_{*}}$ the mass ratio of the planet to the star, \\
\begin{equation}
\label{gamma0}
\Gamma_{0}(r_{P}) = \left(\frac{q}{h}\right)^{2} \, \Sigma(r_{P}) \, r_{P}^{4} \, \Omega_{P}^{2},
\end{equation}
$h=\frac{h_{\mathrm{pr}}(r_{P})}{r_{P}}$,\\
and $\Omega_{P} = \Omega(r_{P})$ the Keplerian angular velocity at the planet position in the disk.

In the isothermal case, $\gamma_{\mathrm{eff}} = 1$, whereas in the adiabatic case, $\gamma_{\mathrm{eff}} = \gamma = 1.4$. Following \citet{paar11}, the effective index is defined by

\begin{equation}
\gamma_{\mathrm{eff}} = \frac{2 Q \gamma}{\gamma Q + \frac{1}{2} \sqrt{2 \sqrt{(\gamma^{2}Q^{2}+1)^{2} + 16Q^{2}(\gamma - 1)} + 2 \gamma^{2}Q^{2} - 2}}
,\end{equation}
where $Q$ accounts for the thermal diffusion:
\begin{equation}
Q = \frac{2 \chi_{P} r_{P}}{3 h_{pr}^{3}(r_{P}) \Omega_{P}}
\end{equation}
and $\chi_{P}$ is the thermal conductivity at the planet location
\begin{equation}
\chi_{p} = \frac{16 \gamma (\gamma - 1) \sigma T_{P}^{4}}{3 \kappa_{P} \rho_{P}^{2} h_{pr}^{2}(r_{P}) \Omega_{P}}
,\end{equation}
with $\sigma$ the Stefan-Boltzmann constant, $\rho_{P}$ the density, $\kappa_{P}$ the Rosseland mean opacity at the planet location, and the 16 factor is a correction introduced by \citet{bitsch11} of a previous 4 factor from \citet{paar11}.

\label{torquecor}
\citet{tanaka02} initially describe a torque arising from the density gradient, called "barotropic" corotation torque. \citet{paarm06} show that the total torque exerted on a planet by the disk increases with the disk opacity and describe how that torque excess is related to the fact that the coorbital region trailing the planet was hotter and underdense. \citep{baruteau08} then show that, in an adiabatic disk, these regions (a colder denser region leading the planet, and a hotter underdense one trailing the planet) could form by entropy advection. The positive resulting "entropic" corotation torque can then become larger than the Lindblad torque.

These two contributions are known to have both linear and non-linear contributions. In the usual range of viscosity ($\alpha_{\mathrm{visc}} < 0.1$), \citet{paarp09b} state that the corotation torques are mostly non-linear, owing to the horseshoe drag caused by the interaction of the planet with the gas in its vicinity \citep{ward91}. As the horseshoe region is closed, it contains a limited amount of angular momentum and therefore is prone to saturation, which cancels the horseshoe contributions to the corotation torque.

\citet{paar10} described the fully unsaturated horseshoe drag expressions for both entropic and barotropic (or vortensity) terms. Using the gravitational softening $b=0.4 h_{\mathrm{pr}}$ also used in \citet{bitsch11} and \citet{bitsch14}, the horseshoe drag torques read:

\begin{eqnarray}
\label{gammahsentro} \gamma_{\mathrm{eff}} \frac{\Gamma_{\mathrm{hs,entro}}}{\Gamma_{0}(r_{P})} &=& \frac{7.9}{\gamma_{\mathrm{eff}}} \, \left(-\frac{\partial \ln T}{\partial \ln r} + (\gamma_{\mathrm{eff}}-1) \frac{\partial \ln \Sigma}{\partial \ln r} \right)_{r_{P}}\\
\label{gammahsbaro} \gamma_{\mathrm{eff}} \frac{\Gamma_{\mathrm{hs,baro}}}{\Gamma_{0}(r_{P})} &=& 1.1 \left(\frac{\partial \ln \Sigma}{\partial \ln r} + \frac{3}{2}\right)_{r_{P}}
.\end{eqnarray}

In the general case (including possibly saturation), the total corotation torque is the sum of the barotropic and entropic contributions:
\begin{equation}
\label{gammacor2}
\Gamma_{\mathrm{corotation}} = \Gamma_{\mathrm{c,baro}} + \Gamma_{\mathrm{c,entro}}
,\end{equation}
each of these contributions including a combination of non-linear (Equations \ref{gammahsentro}-\ref{gammahsbaro}) and linear parts (Equations \ref{gammalinentro}-\ref{gammalinbaro}). The fully unsaturated linear expressions are

\begin{eqnarray}
\label{gammalinentro} \gamma_{\mathrm{eff}} \frac{\Gamma_{\mathrm{lin,entro}}}{\Gamma_{0}(r_{P})} &=& \left(2.2 - \frac{1.4}{\gamma_{\mathrm{eff}}}\right) \, \left(-\frac{\partial \ln T}{\partial \ln r} + (\gamma_{\mathrm{eff}}-1) \frac{\partial \ln \Sigma}{\partial \ln r} \right)_{r_{P}}\\
\label{gammalinbaro} \gamma_{\mathrm{eff}} \frac{\Gamma_{\mathrm{lin,baro}}}{\Gamma_{0}(r_{P})} &=& 0.7 \left(\frac{\partial \ln \Sigma}{\partial \ln r} + \frac{3}{2}\right)_{r_{P}}
.\end{eqnarray}

Thus, \citet{paar11} express the barotropic and entropic torques that account for the saturation effects, as follows:
\begin{eqnarray}
\label{gammacentro}
\Gamma_{\mathrm{c,entro}} &=& F(p_{\nu}) F(p_{\chi}) \sqrt{G(p_{\nu}) G(p_{\chi})} \, \Gamma_{\mathrm{hs,entro}}  \nonumber\\
&& +\,  \sqrt{(1 - K(p_{\nu}))(1 - K(p_{\chi}))} \, \Gamma_{\mathrm{lin,entro}}\\
\label{gammacbaro} \Gamma_{\mathrm{c,baro}} &=& F(p_{\nu}) G(p_{\nu}) \, \Gamma_{\mathrm{hs,baro}} \, + \, (1 - K(p_{\nu})) \, \Gamma_{\mathrm{lin,baro}}
,\end{eqnarray}

where the function $F(p)$ governs saturation:
\begin{equation}
F(p) = \frac{1}{1+(p/1.3)^{2}},
\end{equation}

and the functions $G(p)$ and $K(p)$ govern the cut-off at high viscosity:
\begin{equation}
G(p) = \left\lbrace
\begin{array}{ccc}
\frac{16}{25} \left( \frac{45 \pi}{8}\right)^{3/4} p^{3/2}  & \mbox{for} & p < \sqrt{\frac{8}{45 \pi}}\\
1 - \frac{9}{25} \left( \frac{8}{45 \pi}\right)^{4/3} p^{-8/3} & \mbox{for} & p \geq \sqrt{\frac{8}{45 \pi}}
\end{array}\right.
\end{equation}

\begin{equation}
K(p) = \left\lbrace
\begin{array}{ccc}
\frac{16}{25} \left( \frac{45 \pi}{28}\right)^{3/4} p^{3/2}  & \mbox{for} & p < \sqrt{\frac{28}{45 \pi}}\\
1 - \frac{9}{25} \left( \frac{28}{45 \pi}\right)^{4/3} p^{-8/3} & \mbox{for} & p \geq \sqrt{\frac{28}{45 \pi}}
\end{array}\right.
.\end{equation}
with $p_{\nu}$ the saturation parameter related to viscosity and $p_{\chi}$ the saturation parameter associated with thermal diffusion:
\begin{eqnarray}
p_{\nu} &=& \frac{2}{3} \, \sqrt{\frac{r_{P}^{2}\Omega_{P}x_{s}^{3}}{2 \pi \nu_{P}}}\\
p_{\chi} &=& \sqrt{\frac{r_{P}^{2}\Omega_{P}x_{s}^{3}}{2 \pi \chi_{P}}},
\end{eqnarray}
$\nu_{P}$ the kinematic viscosity at the planet location, $\chi_{P}$ the thermal conductivity at the planet position, and $x_{s}$ the half width of the horseshoe
\begin{equation}
\label{xs}
x_{s} = \frac{1.1}{\gamma_{\mathrm{eff}}^{1/4}} \, {\left( \frac{0.4}{\epsilon / h}\right)}^{1/4} \sqrt{\frac{q}{h}}
\end{equation}
where the smoothing length is $\epsilon / h = b / h_{pr} = 0.4$.

The various contributions of the corotation torque are strongly sensitive to the temperature and surface mass density gradients. Both the corotation and Lindblad torques also scale with $M_{\mathrm{P}}^2$ through $\Gamma_{0}$. The corotation torque also varies with the mass of the planet through the half-width of the horseshoe region.

\label{commentbitsch}\citet{paarp09a} show that the planet mass for which the corotation torque saturates is an increasing function of the viscosity. Actually, their Figure 14 anticipates this mass to grow as $\nu_{\mathrm{visc}}^{2/3}$. Whereas \citet{bitsch15a} use a viscosity of $\alpha_{\mathrm{visc}} = 0.0054$, ours is $\alpha_{\mathrm{visc}} = 0.01$. Therefore, we should expect a slightly larger planet mass where the corotation torque saturates and cannot cancel the Lindblad torque in our case (around 1.5 times greater than in the \citet{bitsch15a} case).

\subsection{Planetary trap evolution}

In this paper, we successively model a variety of planet masses from $1 M_{\mathrm{\oplus}}$ to $200 M_{\mathrm{\oplus}}$ in interaction with an evolving protoplanetary disk, which is located between 0.1 and 30 au, to focus on the planetary formation region. The total torque exerted by the disk over the planet is given by

\begin{equation}
\label{gammatot}
\Gamma_{\mathrm{tot}} = \Gamma_{\mathrm{Lindblad}} + \Gamma_{\mathrm{corotation}}
.\end{equation}

This total torque is positive when the disk yields angular momentum to the planet, making it migrate outward. A negative torque however results in the disk gaining angular moment from the planet, making the planet migrate inward. \citet{lyra10} described the zero-torque interface as an "equilibrium radius". There are two different kinds: places of convergence called "traps" where the inner torque is positive and the outer is negative, and places of divergence, the planetary "deserts" (\citet{hasegawa12} and \citetalias{baillie15}), where the inner torque is negative and the outer positive. Planetary embryos accumulate in traps, while deserts are expected to be depleted in planets.

Assuming fixed surface mass density radial profiles, \citet{hasegawa112} estimated planet trap locations in steady-state disks based on Lindblad torques alone. \citet{bitsch14} proposed that Neptune like planets could get trapped in the regions of outward migration owing to shadowed regions, while \citetalias{baillie15} then studied dynamically evolving disks and showed that the traps are mostly located at the outer edge of the temperature plateaux, and the outer edge of the shadowed regions.

Figure \ref{traptime} shows the evolution of the traps and deserts along the disk dynamical evolution for various planet masses from $1 M_{\mathrm{\oplus}}$ to $200 M_{\mathrm{\oplus}}$. In a typical MMSN-disk, density and temperature gradients are such that the total torque induces inward migration. Therefore, a change of sign of the total torque is generally accompanied by a second one to maintain inward migration in the inner and outer disk regions. Thus, there is always an even number of zero-torque lines with, for each couple, an inner desert and an outer trap.

For a low mass-planet ($1 M_{\mathrm{\oplus}}$), we notice a single "double inversion" located around 10-12 au, at the location of the heat transition barrier between an inner region that is dominated by viscous heating and an outer region being dominated by irradiation heating. \citetalias{baillie15} show that this barrier affects the temperature, density and geometrical profiles of the disk, and that it is a potential planet trap location. However, 2D-disk models from \citet{bitsch14, bitsch15a} do not show evidence of such a strong barrier in the temperature profile and, consequently, do not find any trap in the migration maps around that location. Although their vertical disk profile is initially passive, their simulations involve a proper treatment of the viscous heating. And, although their model accounts for a consistent geometry (including the shadowing effects), we believe that our approach is more self-consistent in relating the disk structure to its evolution: the disk temperature provides the viscosity necessary for calculating the temperature as well as the mass accretion rates across the disk that will, in turn, affect the density profile that constrains back the temperature profile. However, their simulations take into account the heat diffusion, which is not considered in the code presented in this paper and should be in the scope of a future work. This effect should definitely smooth the midplane temperature discontinuities at the transition. In a future paper, we will investigate the question of the sustainability of the heat-transition barrier trap with radial heat diffusion.

Between $5 M_{\mathrm{\oplus}}$ and $20 M_{\mathrm{\oplus}}$, we observe several "double inversions". The outer one is now located a few au inside the heat transition barrier and is now slightly different to the ones observed for $M_{P} = 1 M_{\mathrm{\oplus}}$  and $M_{P} \ge 50 M_{\mathrm{\oplus}}$. \citetalias{baillie15} describe in detail the importance of the temperature plateaux with regard to the positions of traps and deserts. These plateaux, owing to the sublimation of one of the dust's major components, present the most brutal temperature gradient variations, along with the heat-transition barrier that separates the inner disk, which is dominated by viscous heating, and the outer one, which is dominated by irradiation heating. As the torque expressions show, temperature gradients are expected to have a stronger influence on the total torque than the density gradients. Therefore, total torque inversions are correlated to the edges of the temperature plateaux. As \citetalias{baillie15} state, we should refer to sublimation zones rather than lines: in this paper, we assume that the sublimation line just marks the middle of the so-called sublimation zone. Therefore, at the inner edge of this zone, where the temperature gradient is negative inside and almost zero outside, we expect a transition from a positive torque to a negative torque, resulting in a convergence location for planetary embryos (trap). Similarly, a divergence location (planetary desert) is expected at the outer edge of the temperature plateau, where the temperature will resume decreasing. Thus, traps and deserts are expected to be related to sublimation zones. This is consistent with the observation of traps located at the sublimation lines of the water ice, volatile organics, troilite, and between refractory organics and troilite; while deserts are found 1 au outside the water, volatile organics, and refractory organics ice lines, just inside the troilite line and at the silicates' sublimation line. However, we notice that planets more massive than $10 M_{\mathrm{\oplus}}$ are less inclined to be trapped at the volatile organics sublimation line.

In addition, intermediate mass planets ($5 M_{\mathrm{\oplus}}$ to $50 M_{\mathrm{\oplus}}$) can be trapped around 0.3 au. These traps seem to be transient since they appear early in the disk evolution and disappear in less than 10 000 years, a long time before the steady state is reached.

For planets more massive than $20 M_{\mathrm{\oplus}}$, the silicate sublimation line no longer generates deserts after 10 000 years. For more massive planets, traps and deserts associated with the sublimation lines are only possible in the early ages of the disk and progressively disappear completely for masses greater than $100 M_{\mathrm{\oplus}}$. In this case, the only remaining trap and desert are located at the heat transition barrier.

Based on the relative abundances of the dust elements (Table \ref{tempchgt}) and on the opacity jumps associated with the sublimation of these elements (Figure \ref{opa}), we can expect the sublimation of the water ice (the major component), refractory organics, and silicates to affect the medium opacity in the most powerful way, while the opacity will be less sensitive to the troilite sublimation and even less to the volatile organics sublimation. The huge variation in the opacity within a very small range of temperature is reflected in the width of the corresponding temperature plateau, and therefore in the radial separation between the associated trap and desert. The amount of planetary embryos that are expected to accumulate at a given trap grows with the area of convergence towards that trap in the disk, i.e. with that radial separation. The most effective traps should therefore be the water ice trap, the refractory organics trap, and the silicates trap.

\section{Discussion} \label{disc}

\subsection{Influence of the planet mass}
Planetary growth influences the planet dynamics. In this section, we investigate the possibility of trapped planets remaining trapped while they grow. It is therefore necessary to understand how a change in the planetary mass affects the total torque experienced by the planet.

As stated in Equation \ref{gamma0}, the planet mass affects the torque amplitudes through the normalization coefficient $\Gamma_{0}(r_{P})$. It also changes the saturation of the corotation torques and, therefore, the total torque value. Section \ref{mmaps} details migration maps, which display the total torque amplitude as a function of the radial distance to the star and the planet mass. As these migration maps are normalized by $\Gamma_{0}(r_{P})$, the corotation torque saturation remains the only effect from the planet that may affect these maps; the dynamical evolution of the disk itself also obviously affects them as the temperature and density profile vary while the disk ages.

The total torque sign is affected by the variations in temperature gradients (more important than the density-gradient variations) at the sublimation temperature of the dust main components. However, these gradients remain negative around the sublimation regions, which is not necessarily the case around the heat transition barrier between viscous heating and irradiation heating domination, where the temperature gradient may become positive. In the case of a disk that is evolved from a typical MMSN, \citetalias{baillie15} show that, for a $10 M_{\mathrm{\oplus}}$-planet in the inner disk (around the sublimation regions), a negative Lindblad torque was compensated for by a positive corotation torque so that the total torque could become positive itself; whereas near the heat transition, it was the temperature gradient that became positive, therefore reversing the Lindblad and the total torques. For lower mass planets, the mass is too low for the corotation torque to grow and balance the Lindblad torque.

Equations \ref{gammacentro}-\ref{xs} show that the viscosity parameter $p_{\nu}$ increases with the planet mass following $p_{\nu} \propto M_{P}^{3/4}$. \citet{paar11} mentioned that for low viscosity parameters (i.e. small planet masses), the corotation torque could possibly be lower than half its maximum value (corresponding to a stronger viscosity). However, as the planet mass increases, the corotation torque increases as well and is now able to compensate for the Lindblad torque. In addition, we notice from these equations that the corotation torque is more sensitive to the density gradient than the Lindblad torque is. As the planet mass gets significantly higher, the corotation torque drops: this cut-off at high viscosity is reflected by the $G$ and $K$ functions tending to $0$.

This is consistent with what we observe in Figure \ref{traptime}: for low-mass planets (e.g. $M_{P} = 1 M_{\mathrm{\oplus}}$) or too massive planets (e.g. $M_{P} \ge 100 M_{\mathrm{\oplus}}$), the corotation torque does not allow the sign of the total torque to reverse and thus traps can only be found at the heat transition where the temperature gradient inversion alone is sufficient to reverse the Lindblad torque just outside the shadow region that is located around this transition.

As the torque exerted by the disk on a planet located in a trap is null, it is very likely that a trapped planet will remain trapped if the trap survives and the planet mass does not vary much. In that case, Figure \ref{traptime} shows that the planets slowly migrate towards the interior of the disk. However, for planetary masses $M_{P} \le 50 M_{\mathrm{\oplus}}$ located beyond 1 au, there can be several inner traps that would mean an escaping planet is trapped again.

\subsection{Type II migration}
The planet mass also favors the opening of a gas gap in the protoplanetary disk. In their Equation 15, \citet{crida06} derived a criterion for opening a gap:

\begin{equation}
\label{eq15crida06}
\frac{3 h_{pr}}{4 R_{Hill}} + \frac{50}{q \mathcal{R}} \lesssim 1
,\end{equation}

where $R_{Hill}$ is the Hill radius ($R_{Hill} = r_{P} \left( \frac{M_{P}}{M_{*}}\right)^{1/3}$) and the Reynolds number is $\mathcal{R} = \frac{r_{P}^{2} \Omega_{P}}{\nu_{P}}$.

The first term varies according to $(h/r)M_{P}^{-1/3}$ and the second one in $(h/r)^{2}M_{P}^{-1}$. Therefore, either an increase in the planet mass or a decrease in the local aspect ratio $h/r$ may favor a gap opening. The aspect ratio profile from \citetalias{baillie15}, Figure 6 showed that the aspect ratio minima are located around 15 au in young disks and may get closer to almost 7 au when the disk ages. Thus, these are places where gaps are most likely to appear in our model if there were planets located in those locations.

In this paper, we consider the action of the disk over a putative planet without taking into account the back-reaction of the planet on the disk. In addition, we chose to consider a simplistic bimodal model for the disk-planet interaction: either the planet and the disk satisfy the criterion from equation \ref{eq15crida06} for opening a gap and the planet changes from type I to type II migration, or we consider it to be still embedded in a gas disk, not opening any gap, experiencing the torques defined by \citet{paar11}, and following type I migration. These torque formulas have been obtained for low-mass planets (for which the horseshoe width has to be lower than the disk scale-height). In our case, the aspect ratio profiles suggest that these torque formulas are valid for a planet up to a few tens of Earth-masses. Though our study extends beyond this mass limit, we consider these expressions as providing a first approximation of the torques experienced by a massive planet in our bimodal migration model. This could be improved by considering the partial opening of gaps in the disk by planets that are more massive than a few tens of Earth masses, by introducing a transition torque similar to \citet{dittkrist14} for example. Additionally, modeling the gap opening realistically would require taking into account the planet back-reaction on the disk, which could be done in the scope of a future work by coupling our hydrodynamical code with a planetary growth code.

\subsection{Migration maps}
\label{mmaps}
In this section, we focus on the torque's dependency on the planet mass and its distance to the star. To allow the torques to be compared, we display normalized torques by $\Gamma_{0}$ (Equation \ref{gamma0}).

\begin{figure}[tp!]
\begin{center} $
\begin{array}{c}
\includegraphics[width=8cm, clip=true]{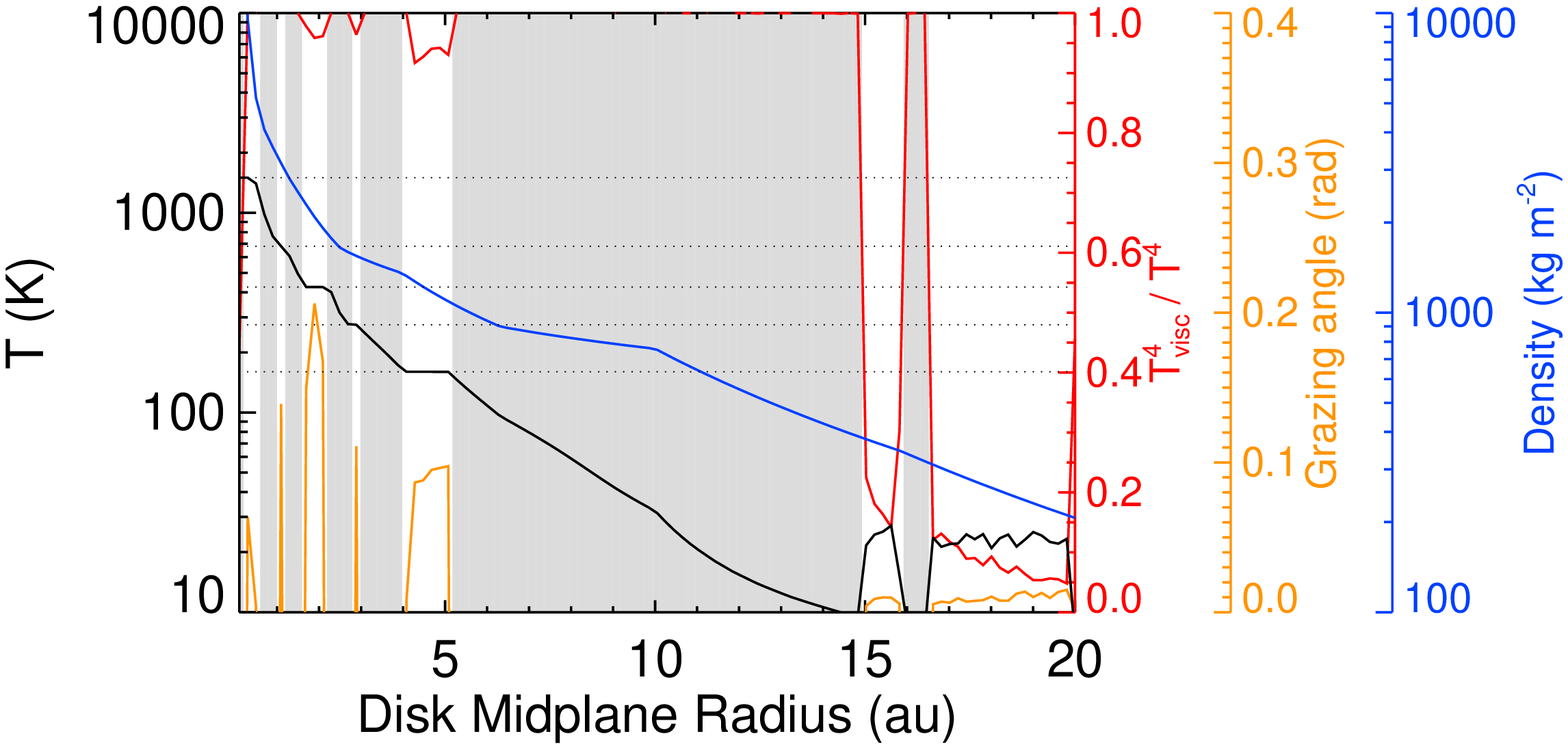}\\
\includegraphics[width=8cm, clip=true]{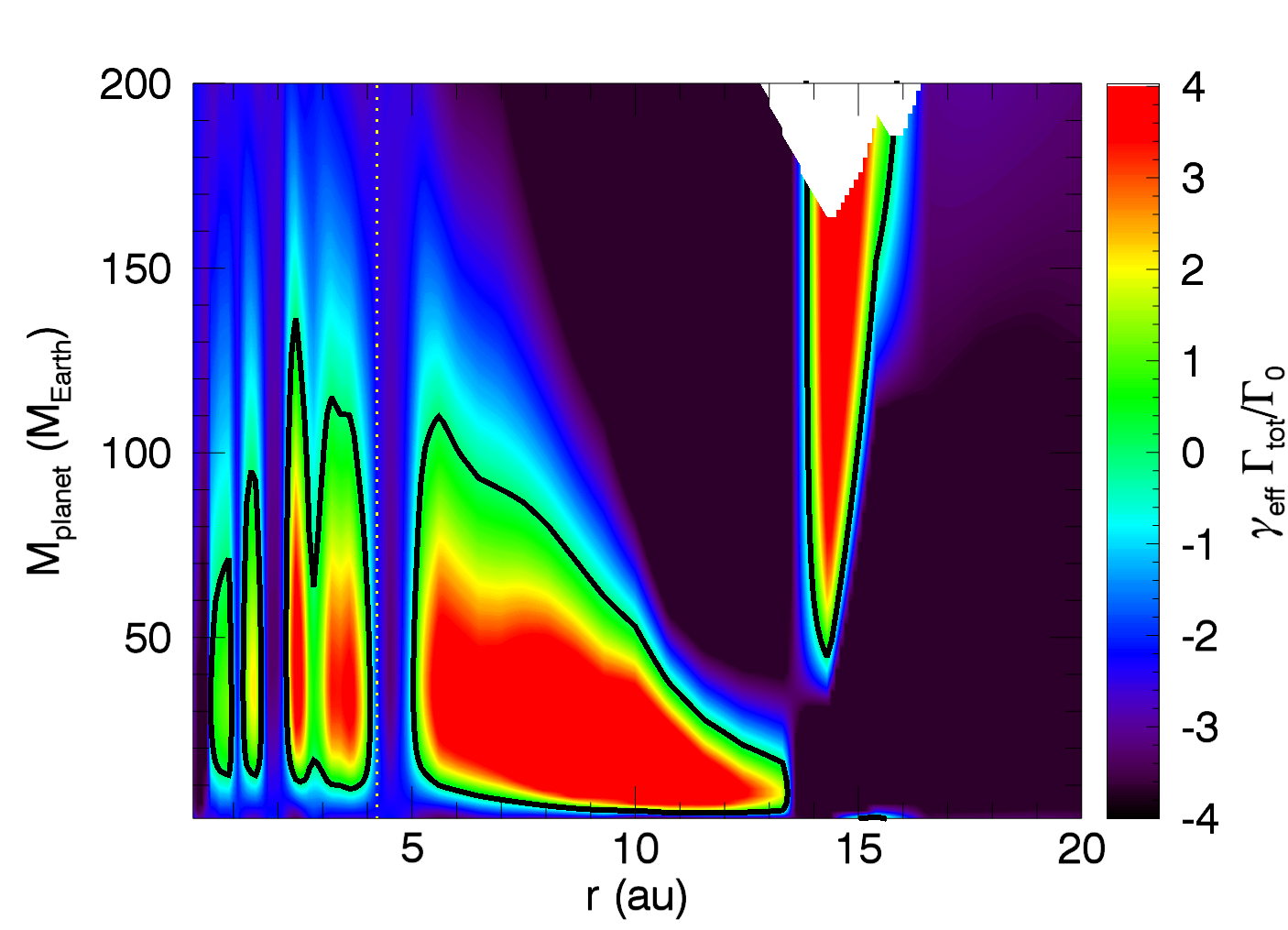}\\
\includegraphics[width=8cm, clip=true]{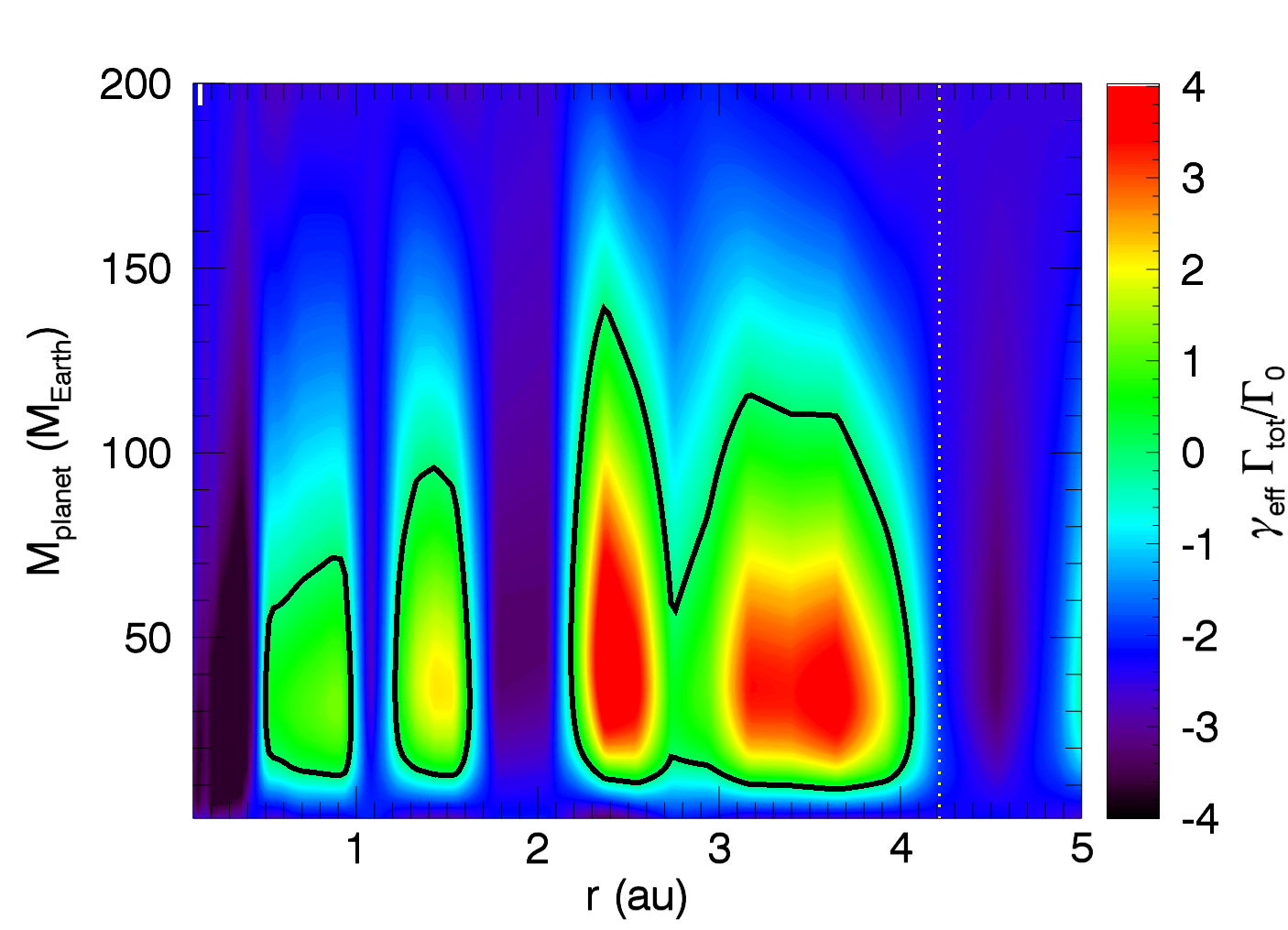}\\
\end{array} $
\end{center}
\caption{Upper panel: Mid-plane temperature (black), surface-mass density (blue), grazing angle (yellow) and viscous heating rate (red) radial profiles after 10,000 years of evolution. Shadowed regions are displayed in gray and sublimation lines are shown in dotted lines. Lower panels: Migration torque of a protoplanet with given radial distance to the central star $r_{P}$ and mass $M_{P}$, in a protoplanetary disk after 10,000 years of evolution. The white area verifies the criterion from Equation \ref{eq15crida06}. Black contours (0-torque contour) delimit the outward migration conditions while the rest of the migration map shows inward migration. Planetary traps are located at the outer edges of the black contours while planetary deserts are at the inner edges. The yellow dotted line marks the water ice line.}
\label{mmap10000}
\end{figure}

\begin{figure}[tp!]
\begin{center} $
\begin{array}{c}
\includegraphics[width=8cm, clip=true]{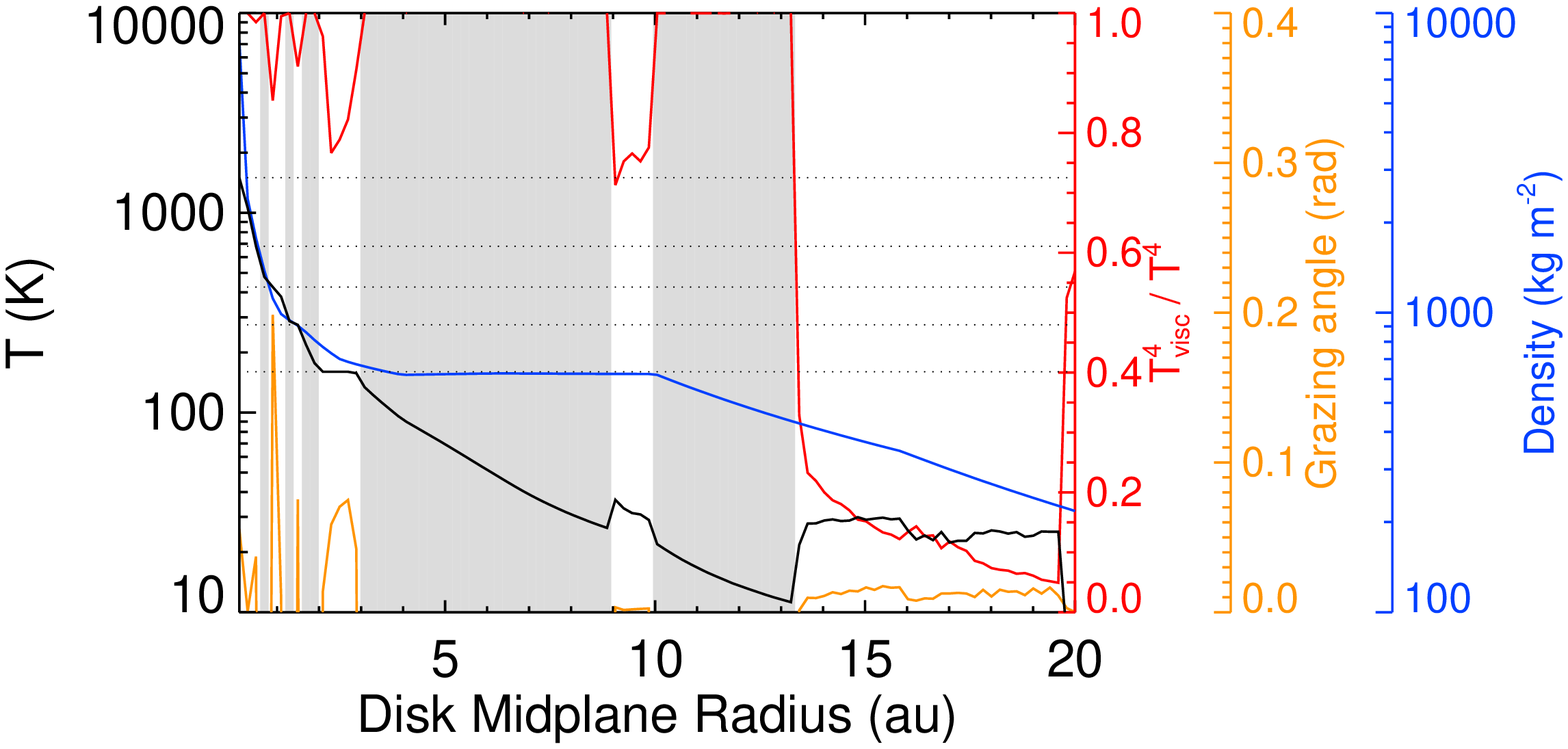}\\
\includegraphics[width=8cm, clip=true]{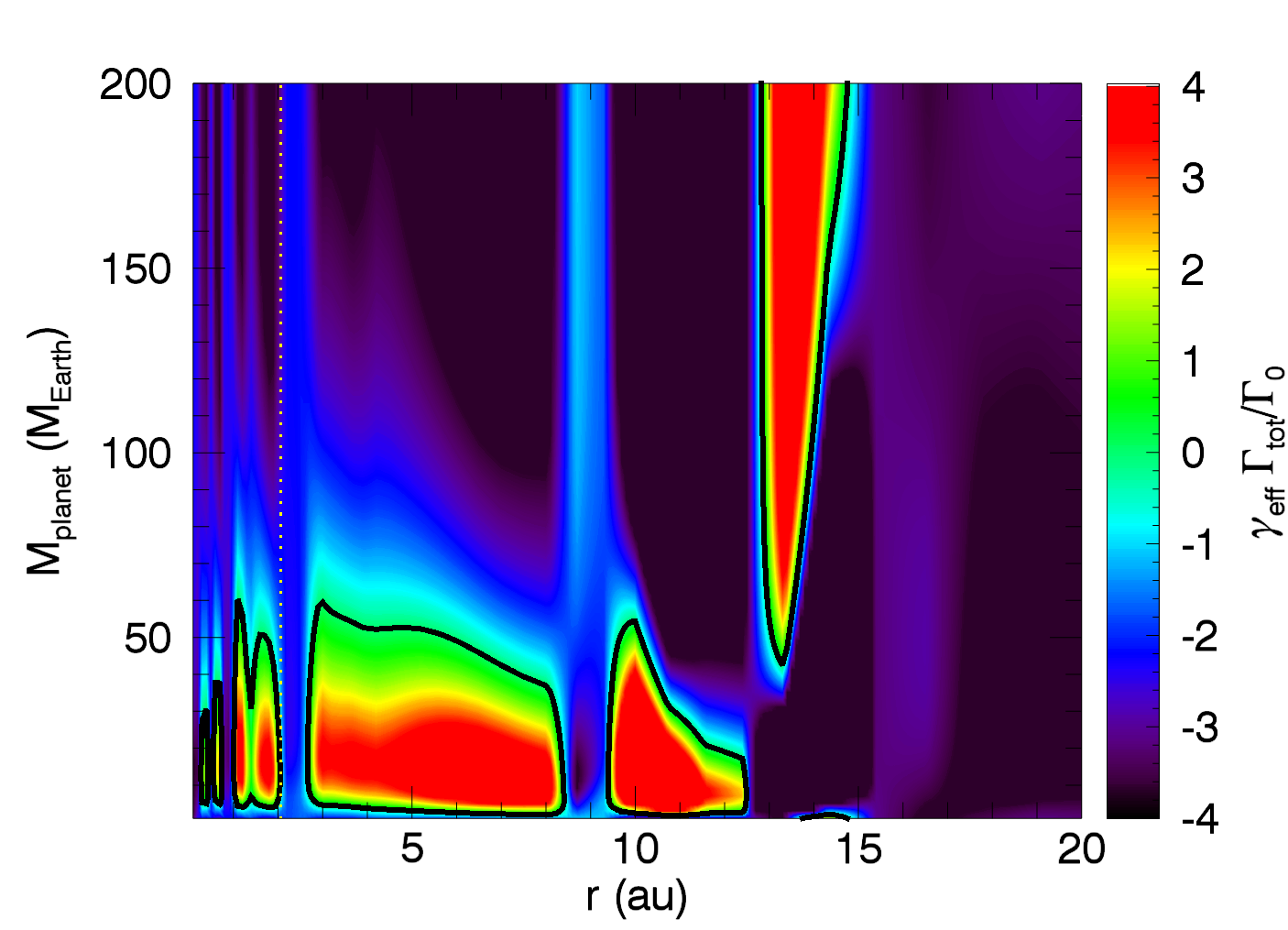}\\
\includegraphics[width=8cm, clip=true]{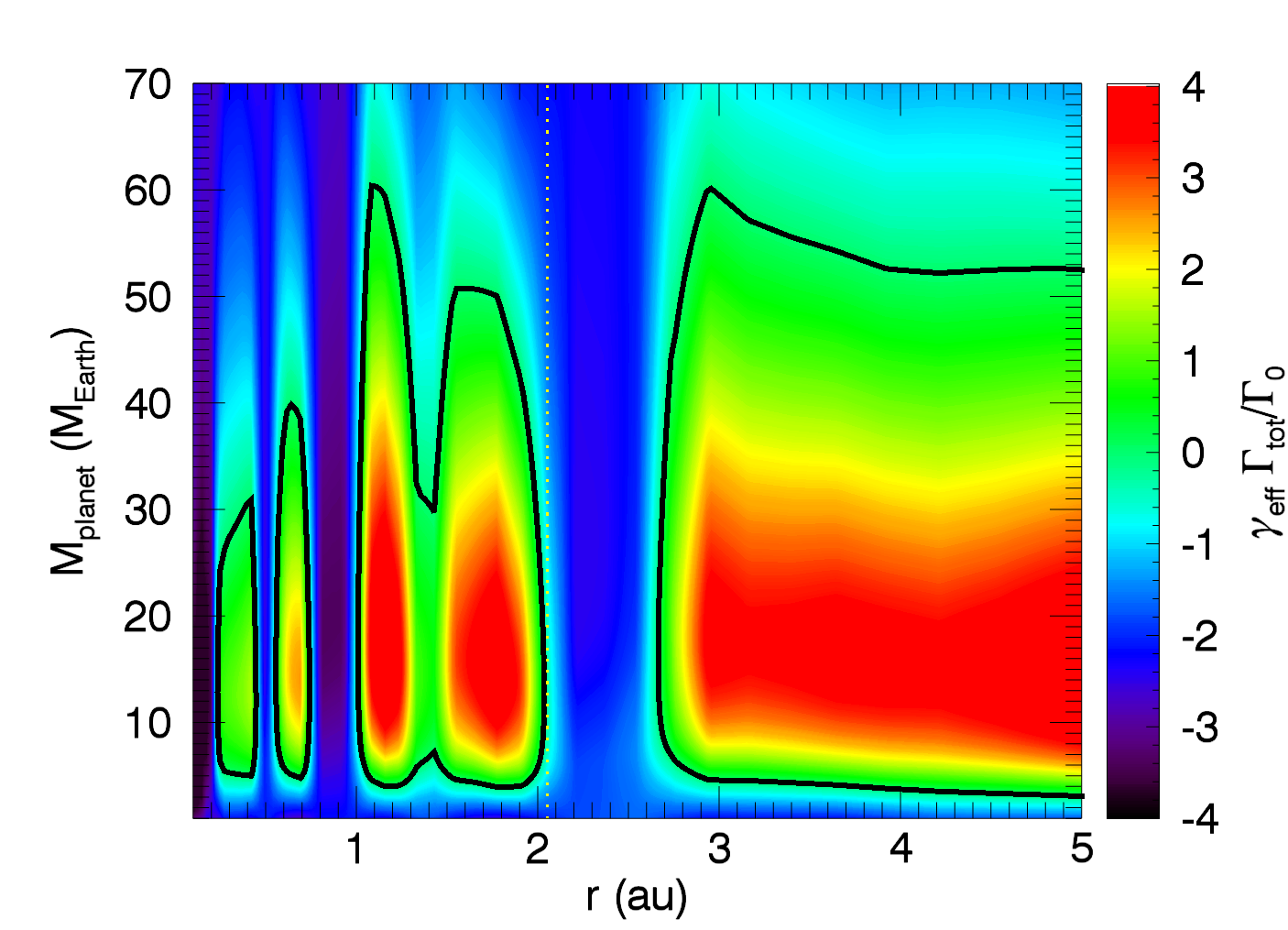}\\
\end{array} $
\end{center}
\caption{Disk profiles and migration maps for a disk after 100 000 years of evolution. Legend is the same as in Figure \ref{mmap10000}.}
\label{mmap100000}
\end{figure}

\begin{figure}[tp!]
\begin{center} $
\begin{array}{c}
\includegraphics[width=8cm, clip=true]{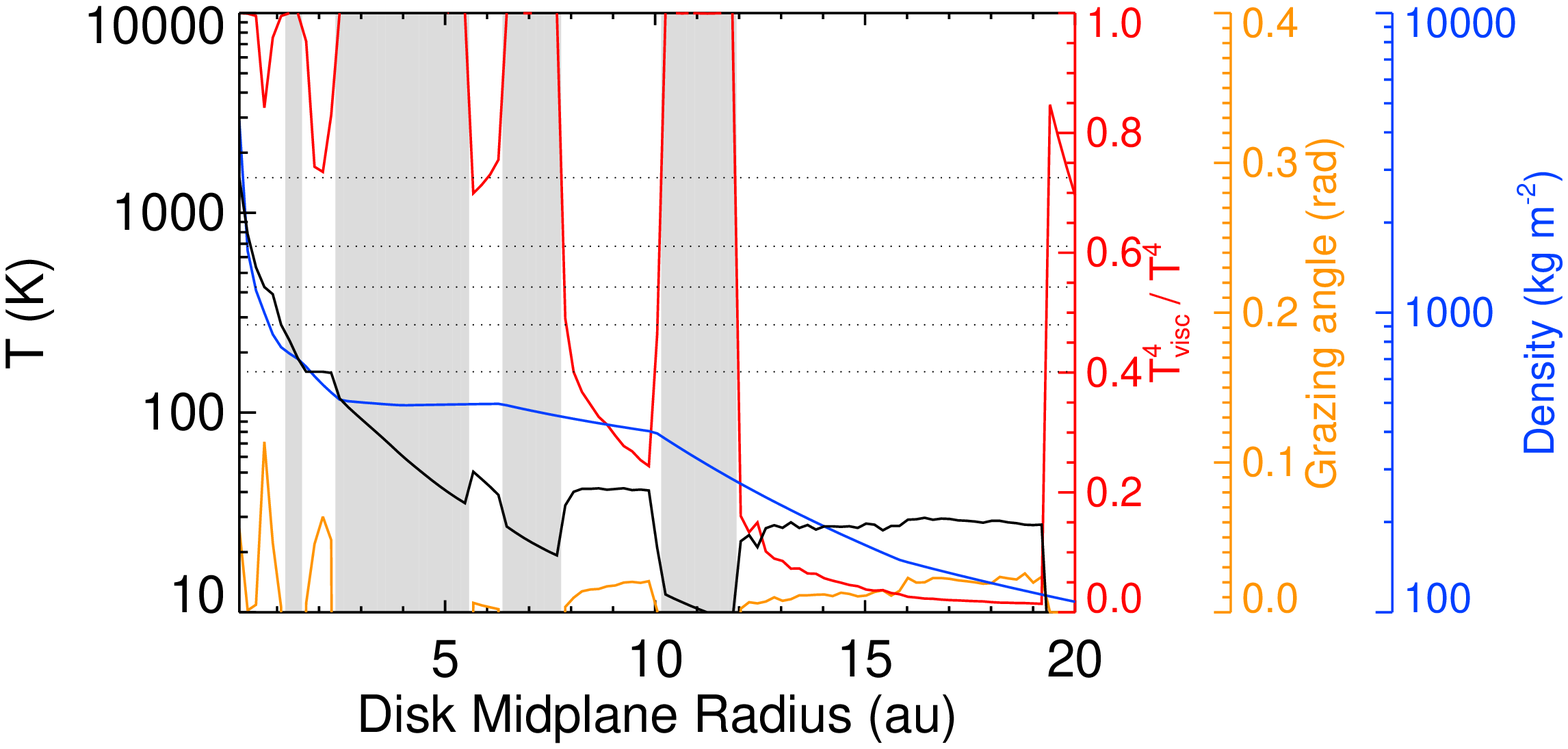}\\
\includegraphics[width=8cm, clip=true]{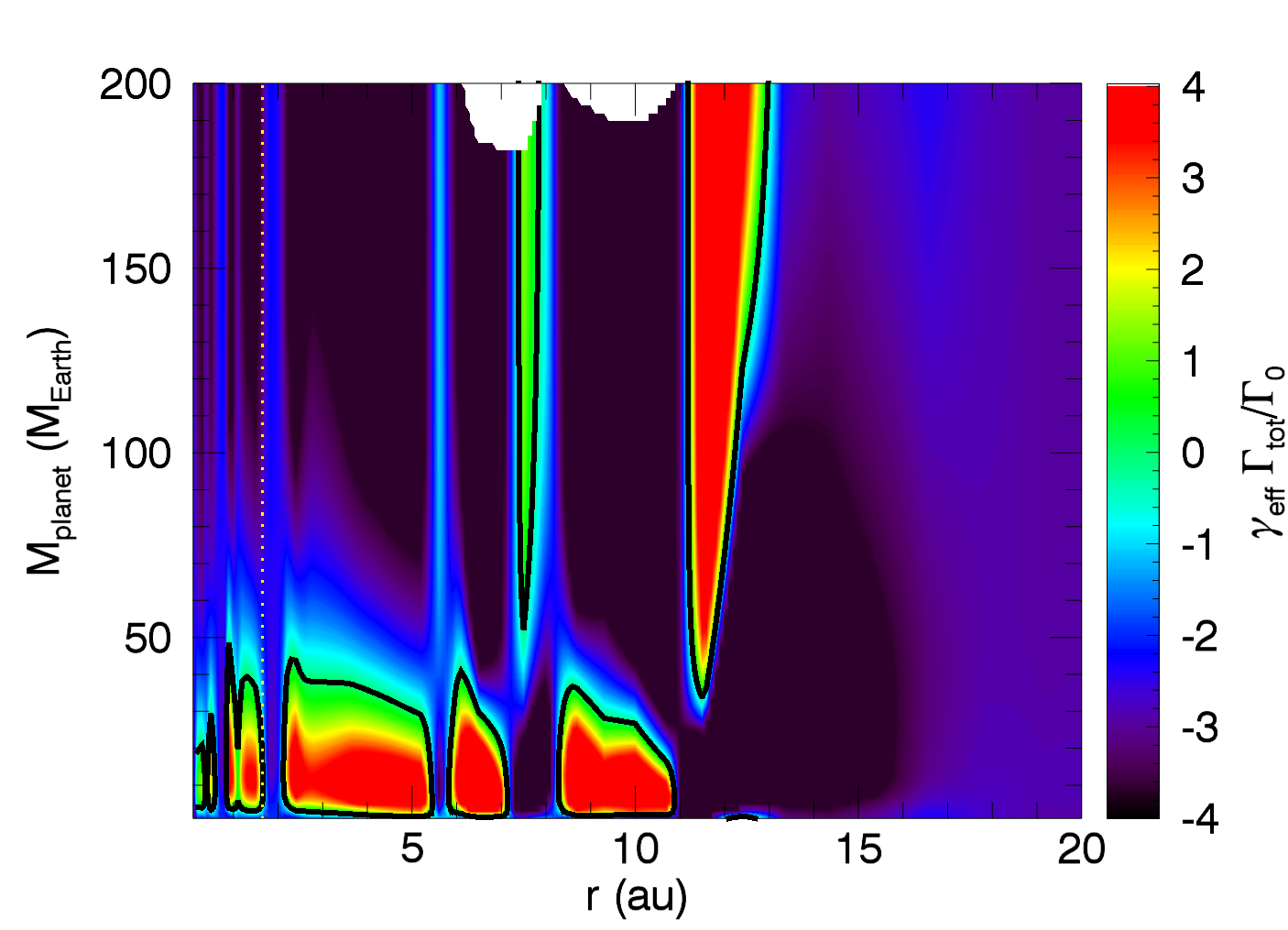}\\
\includegraphics[width=8cm, clip=true]{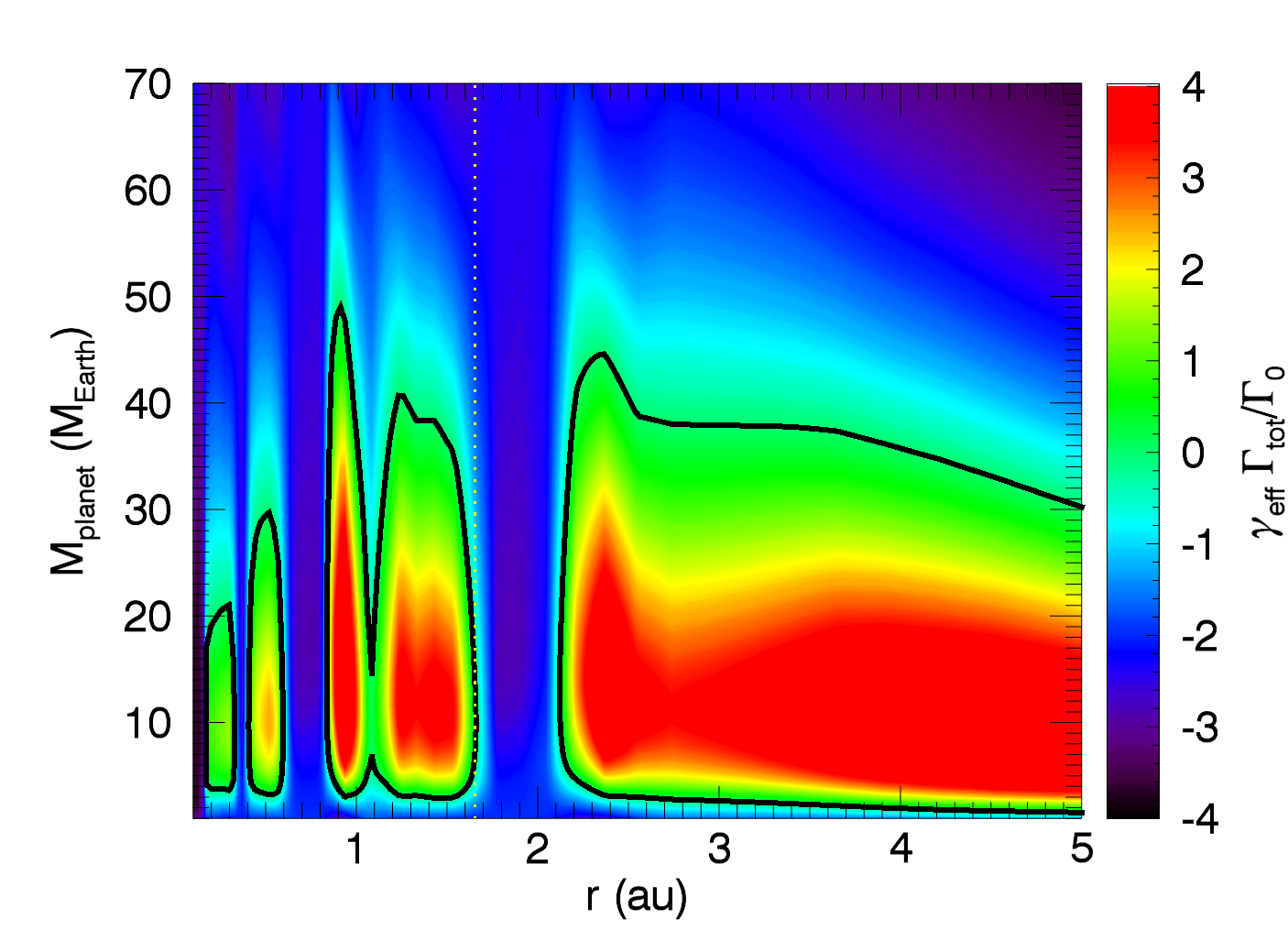}\\
\end{array} $
\end{center}
\caption{Disk profiles and migration maps for a disk after 1 million years of evolution. Legend is the same as in Figure \ref{mmap10000}.}
\label{mmap1000000}
\end{figure}

\begin{figure}[tp!]
\begin{center} $
\begin{array}{c}
\includegraphics[width=8cm, clip=true]{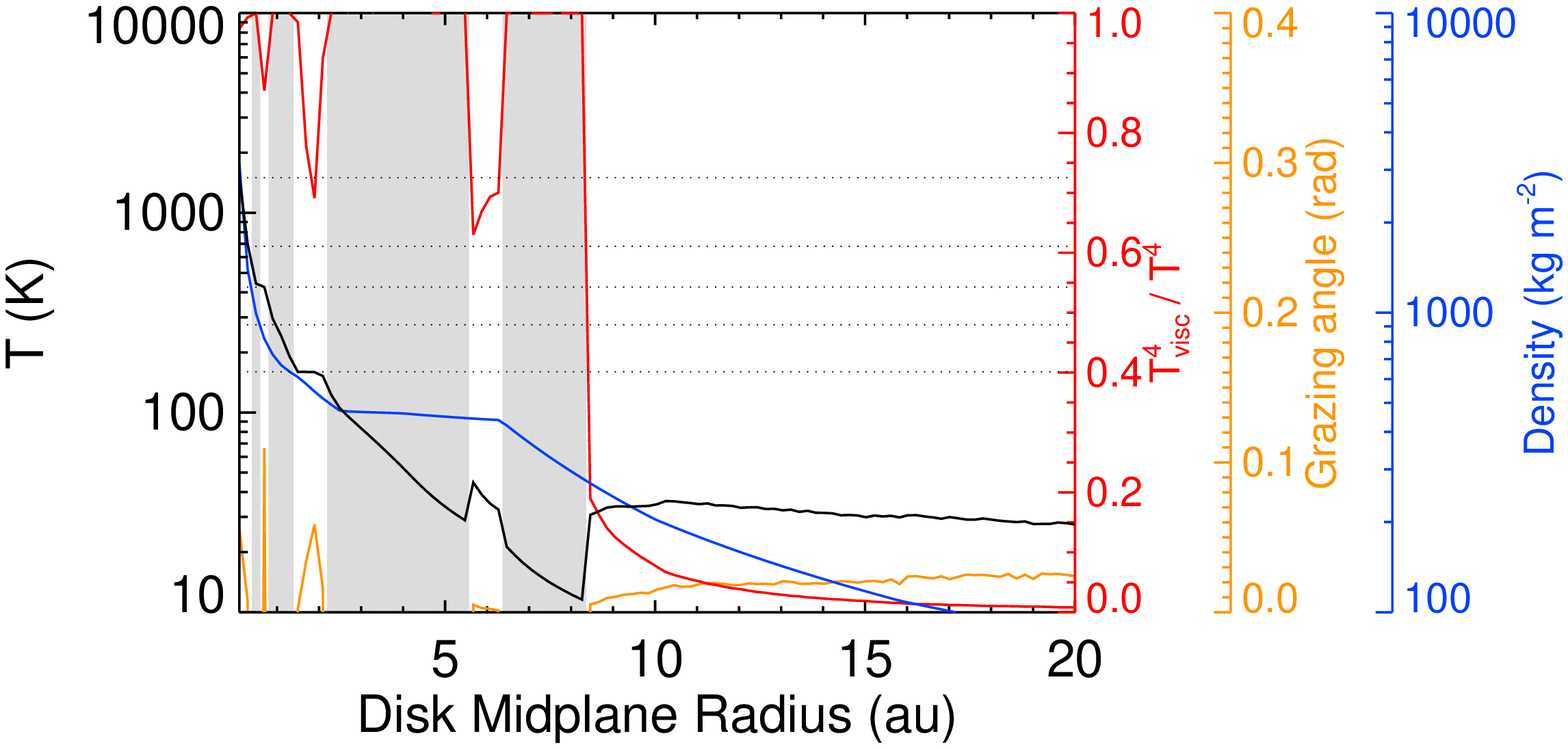}\\
\includegraphics[width=8cm, clip=true]{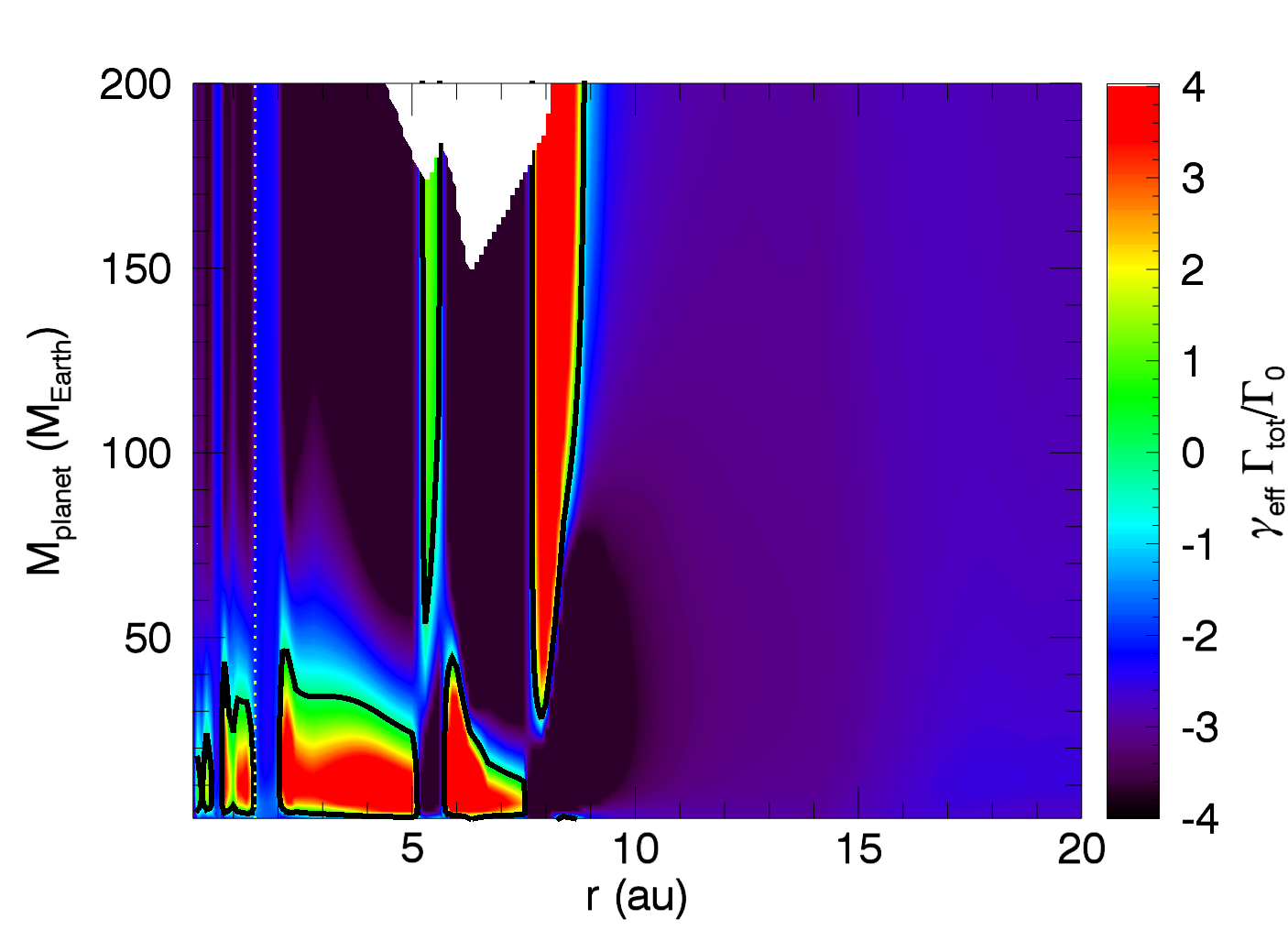}\\
\includegraphics[width=8cm, clip=true]{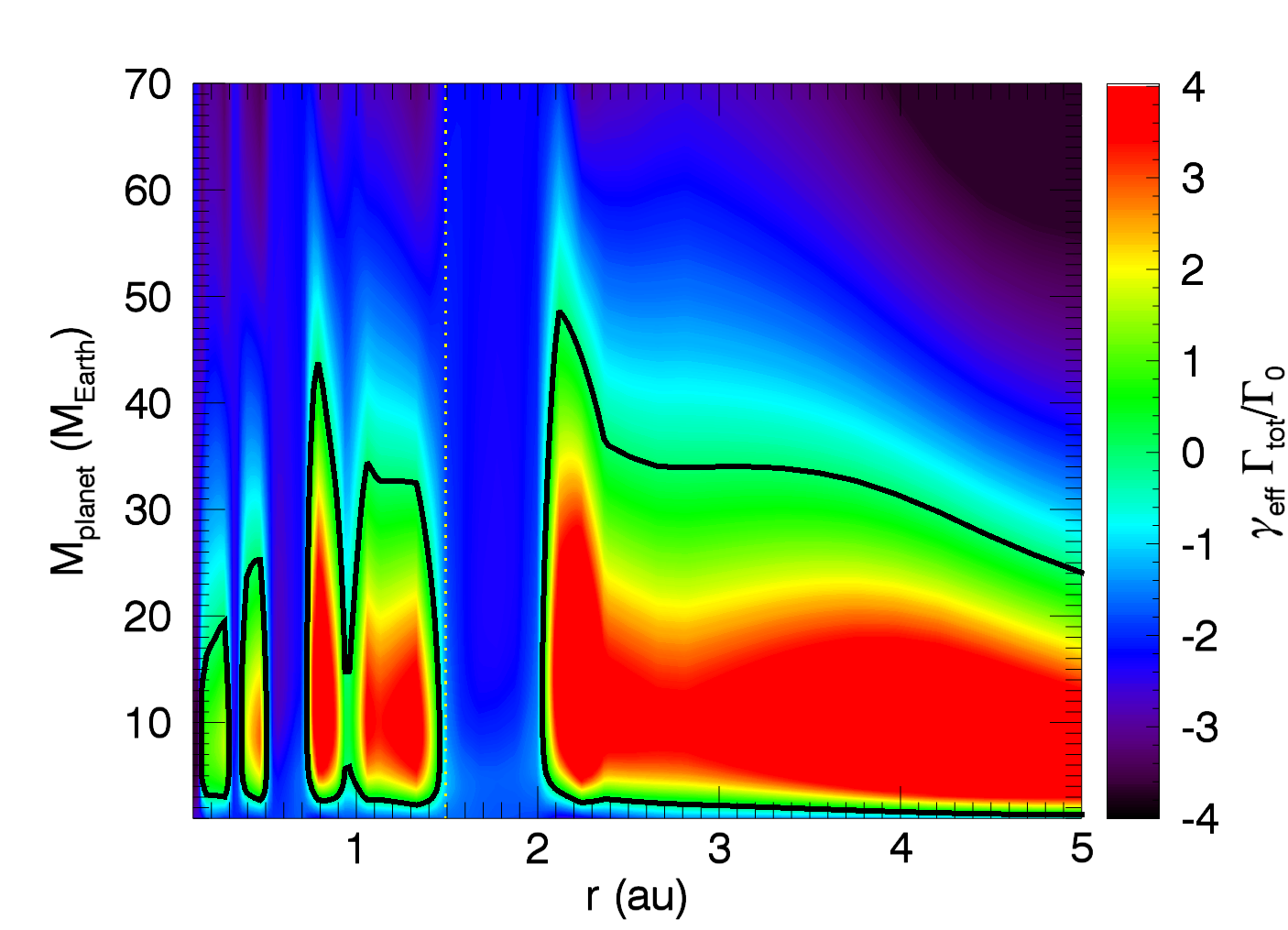}\\
\end{array} $
\end{center}
\caption{Disk profiles and migration maps for a disk after 5 million years of evolution. Legend is the same as in Figure \ref{mmap10000}.}
\label{mmap5000000}
\end{figure}

\begin{figure}[tp!]
\begin{center} $
\begin{array}{c}
\includegraphics[width=8cm, clip=true]{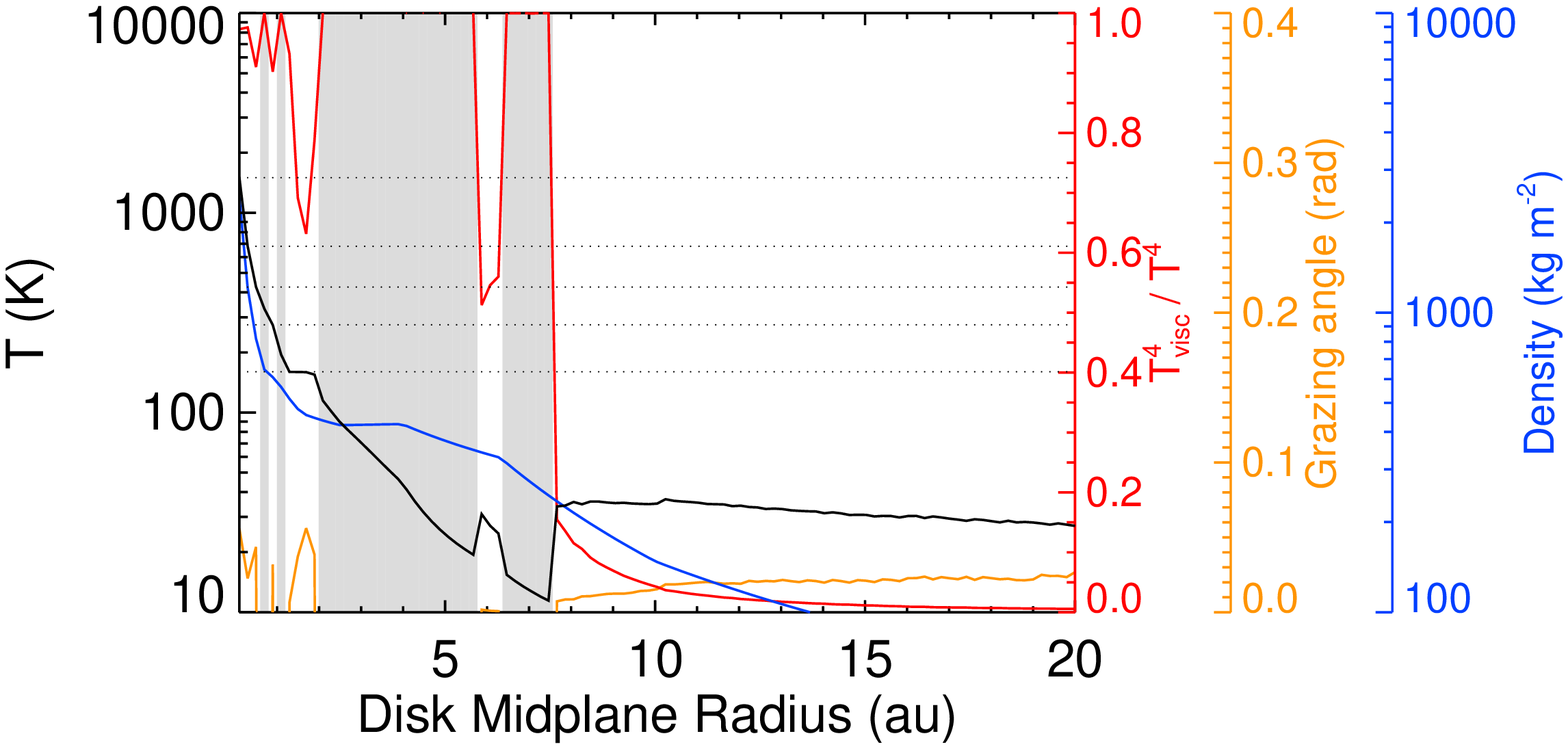}\\
\includegraphics[width=8cm, clip=true]{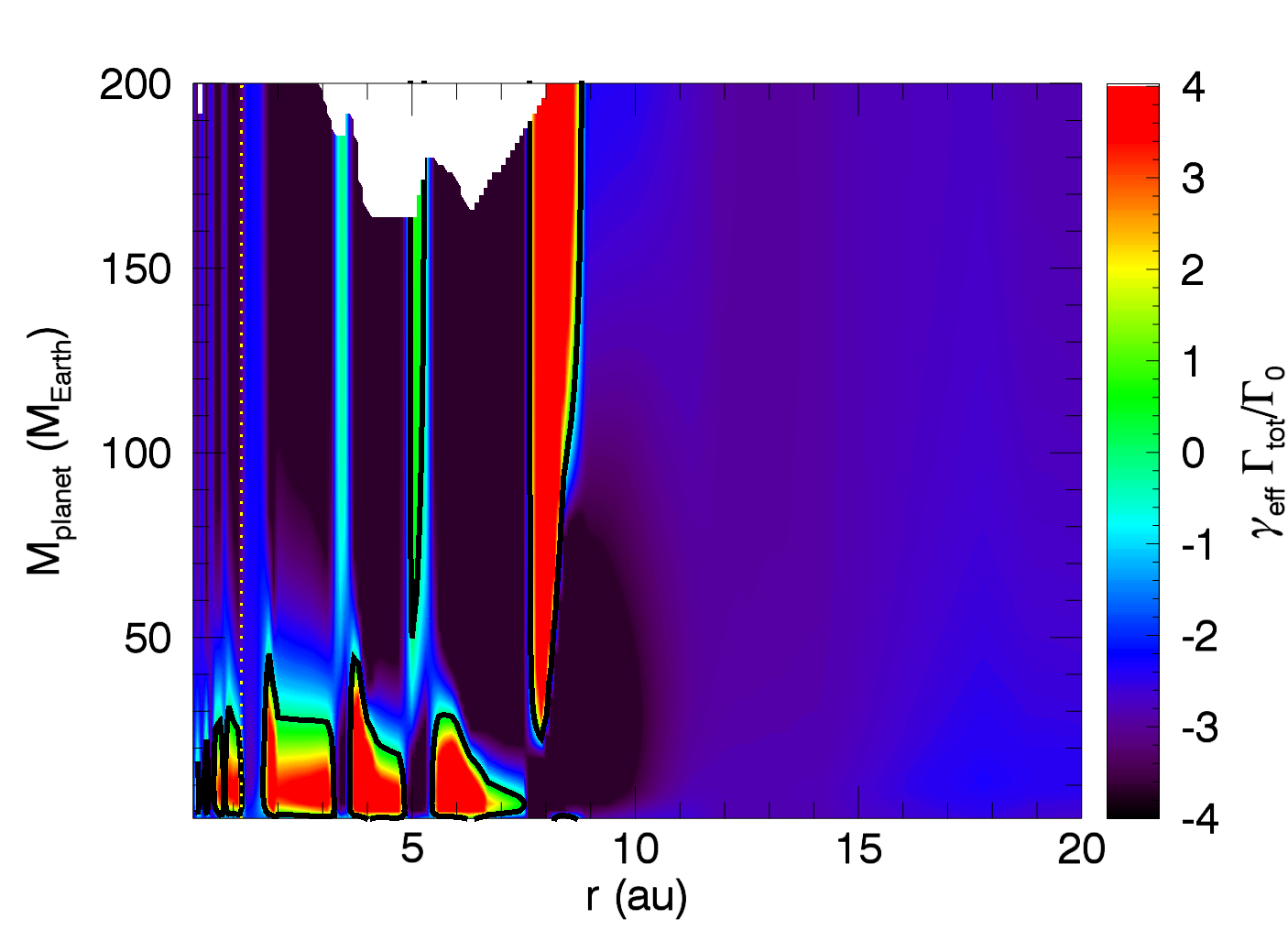}\\
\includegraphics[width=8cm, clip=true]{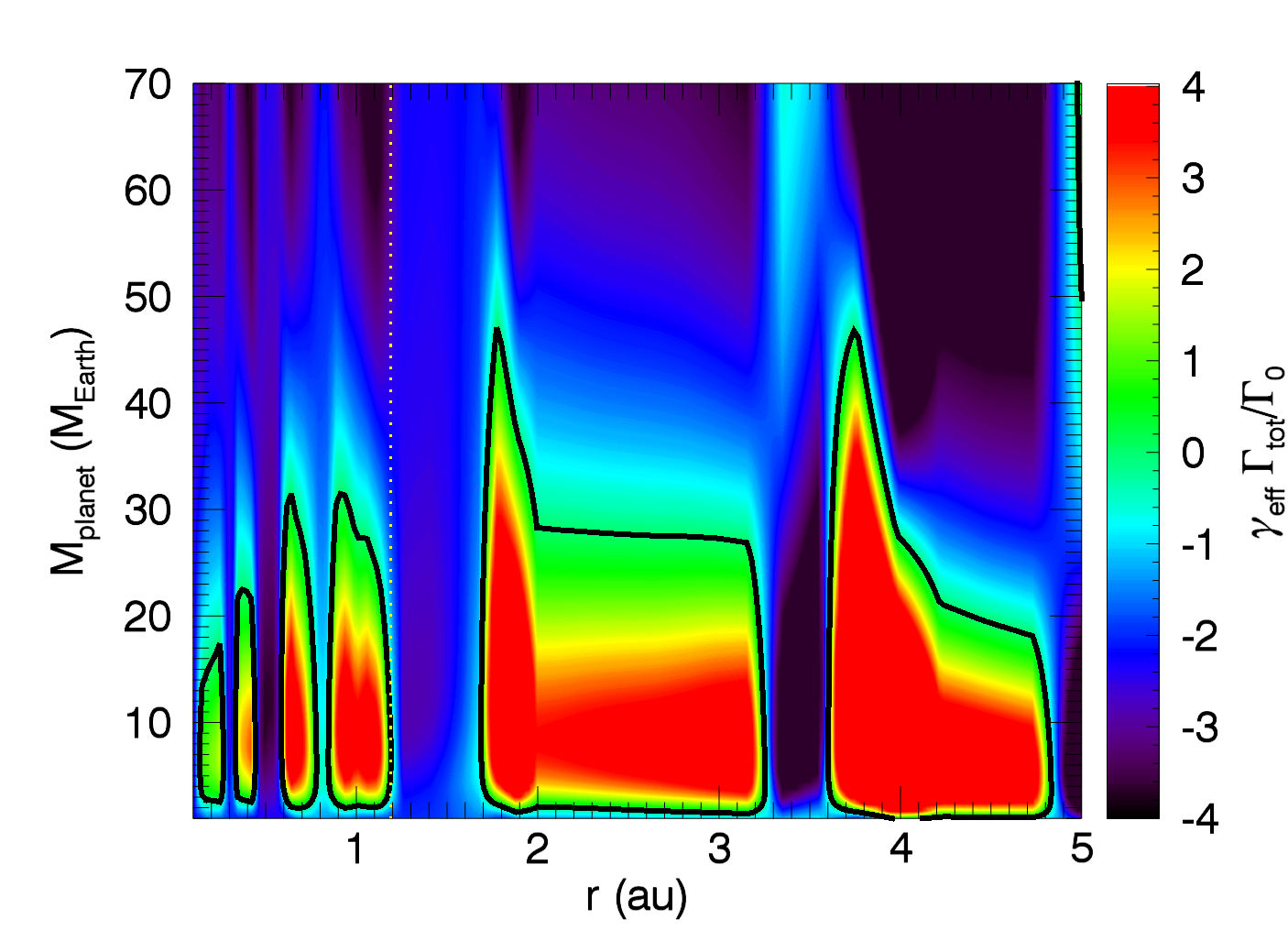}\\
\end{array} $
\end{center}
\caption{Disk profiles and migration maps for a disk after 10 million years of evolution. Legend is the same as in Figure \ref{mmap10000}.}
\label{mmap10000000}
\end{figure}

Figures \ref{mmap10000}-\ref{mmap10000000} show the disk density and temperature profiles in parallel with the torque amplitudes at different radial and planet mass scales. In this distance-mass representation, we notice closed zones of outward migration in a continuum of inward migration. The contours of these zones correspond to the location where the total torque cancels. The outer borders are the planetary traps, while the inner ones are the planet deserts. As seen in the previous section, sublimation lines can be associated with the locations of the planet traps.

We note that there is an actual possibility for a $20 M_{\mathrm{\oplus}}$-planet to get trapped at 12.5 au. And we also note that the outward migration zones correspond to the shadowed regions of the disk as indicated in gray in the upper panel of Figures \ref{mmap10000}-\ref{mmap10000000}.

These outward migration zones associated with the ice lines are wider for lower-mass planets: as the planet mass increases, the trap is found further inside the disk, while the deserts remain at the same location. While we can see from Figures \ref{mmap10000}-\ref{mmap10000000} that these outward migration regions are actually shadowed, BC15's Figure 6 shows that the aspect ratio is locally decreasing across those regions. Equation \ref{xs} states that a decrease in the aspect ratio has a similar effect as a mass increase on the corotation torque saturation. Therefore the corotation torque saturates for lower planet mass in the outer region of the outward migration zone. However, for a disk of a given age, the maximal mass of the planets that can be trapped in the different traps is similar.

The most exterior outward migration zone is also the one that may trap the most massive planets. Though it is not associated with a sublimation line, it is related to the temperature and density profile irregularities that are generated by the heat transition region, as is visible in the upper panel. Furthermore, we note that below $50 M_{\mathrm{\oplus}}$, the outermost trap is no longer related to the heat transition barrier, unlike for more massive planets: as we consider a lower planet mass, the corotation torque is no longer sustained and its saturation prevents the compensation of the Lindblad torque.

In addition, we note that there are zones in the migration maps (white areas) where the planets are massive enough and the aspect ratio low enough for a gap to open: the planet is now in type II migration. After 10 000 years of evolution, planets more massive than $170 M_{\mathrm{\oplus}}$ are able to open gaps around 15 au. Moreover, it appears that any planet located outside the most inner desert can escape type I inward migration: they migrate toward a trap and therefore avoid falling onto the central star. Though we consider the influence of the disk on putative planets of various masses, we do not actually follow their growth, migration, nor back-reaction over the disk.

After 100 000 years, it is no longer possible to trap planets more massive than $60 M_{\mathrm{\oplus}}$ at the sublimation lines. That maximal mass is comparable with that of $38 M_{\mathrm{\oplus}}$ from \citet{bitsch15a}, Figure 4, which was obtained for an accretion rate of $3.5 \times 10^{-8} \, M_{\odot}/yr$, and which is compatible with a disk that is slightly younger than 100 000 years in our case. This maximal mass ratio is consistent with the 1.5 ratio expected in Section \ref{commentbitsch}. We note that the outward migration zone beyond the water ice line for the low-mass planets is now split into two different outward migration zones: we explain the separation between the two zones by the presence of a shadowed region in the disk, which induces a change in the dominant heating processes in this place. In addition, the type II migration area disappears as the aspect ratio is higher at 100 000 years than it was after 10 000 years. It would require a more massive planet ($M_{P} \ge 200 M_{\mathrm{\oplus}}$) to open a gap at this location.

\citetalias{baillie15} indicate that after 1 million years of evolution, the minimal value of the aspect ratio was reached much further into the disk (around 7 au) than it was previously. This generates the white areas in Figures \ref{mmap1000000}-\ref{mmap10000000} where potential planets would open gaps in the disk.

After that date, it becomes very difficult for a planet that is more massive than $40 M_{\mathrm{\oplus}}$ to get trapped at a sublimation line. This upper limit decreases with time. \citetalias{baillie15} shows that, at a given radius, the pressure scale height decreases with time. The half-width of the horseshoe $x_{s}$ then increases, leading to an increase in the viscosity parameter $p_{\nu}$, which in turn induces a stronger saturation. Therefore, the planet mass at which the corotation torque cannot compensate the Lindblad torque decreases.

We also note the apparition of a few new outward migration zones that are not related to the sublimation lines: the increase in the number of shadowed regions leads to an alternation of dominant heating processes, which generate numerous consecutive outward migration zones for the most massive planets. This then helps to trap these massive planets ($M_{P} \ge 50 M_{\mathrm{\oplus}}$) at locations further away from the sublimation lines.

These phenomena are even more obvious after 10 million years (Figure \ref{mmap10000000}). With time, the maximal mass of a planet that can be trapped decreases. This makes the trapping process more and more difficult as the planet mass grows over time.

\section{Trapping growing planets}
\label{trappinggrowingplanets}
From Figures \ref{mmap10000} to \ref{mmap10000000}, we can infer that a trapped planet will remain trapped as it grows. Indeed, for a $60 M_{\mathrm{\oplus}}$-planet trapped at 10 au after 10 000 years, if it were to grow, this planet will enter the inward migration zone that would bring it back inside the trap located just inside its initial position. Similarly, a $100 M_{\mathrm{\oplus}}$-planet that is initially located at 15 au would enter the outward migration region if it grew. It would then migrate outward back to the trap, just outside its previous location. Thus, trapped planets will remain trapped and will follow the trap migration as they gain mass. In most cases, this involves a slow inward migration. If the planets reach the upper mass limit of the outward migration region where they are, gaining mass may free them and they will resume migrating inward. They may be trapped again when they cross the next inner outward migration region if their mass is still lower than the mass limit of that zone. Otherwise, they may not be able to escape the fatal inward migration. The planet growth speed is then the key parameter when estimating whether it will become massive enough to open a gap quickly or if it will not grow fast enough to avoid resuming a fatal inward migration. Since massive planets are already inclined to enter the runaway gas accretion phase, it is highly unlikely that they will remain trapped in the inner traps with a mass not exceeding $100 M_{\mathrm{\oplus}}$). These types of planets would be saved if they do not pass that mass limit before photo-evaporation can dissipate enough gas to slow the planet type I migration.

Though a gap can be opened by a massive enough planet anywhere in the disk, we can infer a plausible scenario for gap-opening with a trapped minimal mass planet: a massive planet that follows a trap is likely to open a gap and enter type II migration at the radial distance at which the trap verifies the type II migration criterion from Equation \ref{eq15crida06} (i.e. at the junction between the black trap contour and the white area in Figures \ref{mmap10000}-\ref{mmap10000000}). After 10 000 years, this could only happen around 15.8 au for planets more massive than $180 M_{\mathrm{\oplus}}$. Similarly, after 1 million years, this can happen at 7.7 au for planets around $180 M_{\mathrm{\oplus}}$. Even though it is very likely that the gas will be dissipated by photo-evaporation at longer evolution times, gaps could appear around 5 au for planets slightly more massive than $170 M_{\mathrm{\oplus}}$.

Thus, we can list several types of planets that may escape type I migration: those that will grow very quickly and open a gap (with masses higher than $200 M_{\mathrm{\oplus}}$), and those that will grow slowly and stay trapped until the gas dissipates and possibly avoid spiraling inward onto the star. The former may open gaps without necessarily experiencing type II migration rates: \citet{durmann15} show that in a disk where mass exceeds approximately $0.2 M_{J}$, a Jupiter mass planet would migrate inward faster than if it was migrating at the viscous rate of the disk, and that the actual migration rate would depend on the disk surface mass density. Therefore, planets forming in the inner disk are still prone to fall onto the star: this is consistent with the results of planetary growth by pebble accretion from \citet{bitsch15b}, which states that gas giants mainly form beyond 10 au (and up to 50 au if they form very early in the disk evolution) before migrating inward over several au. Therefore, investigating the survival of the giant planets requires a more complex model than the type I migration maps we present in this paper. The latter can be found at any distance from the star, although they can exceed $100 M_{\mathrm{\oplus}}$ only if they get trapped in the outermost traps (those not associated with the sublimation lines). In addition, it seems possible to find planets trapped below 1 au.

\section{Super-Earths}
\label{se}

Exoplanet observations mentioned in \citet{schneider11} and \citet{wright11} show the existence of a population of planets with masses ranging between a few Earth masses and a few tens of Earth masses, and located inside 1 au from the star. A certain proportion of these super-Earths appear to be in mutual resonances (see Figure 4a from \citet{ogihara15}). \citet{hansen12} and \citet{hansen13} aimed at modeling the in situ formation of hot-Neptunes and super-Earths inside 1 au, using N-body numerical simulations. However, their simulations required a much more massive solid disk than the MMSN solid part. Although adding gas to such a massive disk would probably be gravitationally unstable, a more realistic disk model would consider a smaller quantity of gas by enriching the inner disk in solids by inward migration of dust grains, pebbles, and planetesimals. Their model however did not manage to reproduce the distribution of period ratios between adjacent planets.

Performing similar but more complete simulations, \citet{ogihara15} modeled super-Earths in situ formation starting with planetary embryos and planetesimals. They included planet-disk interactions, although with a simplistic disk model that assumes the temperature profile of an optically thin disk. They show that the interaction between the planets and the gas disk, and the resulting migration, should not be neglected. However, given their simplistic temperature profile, their disk cannot account for shadowing effects, temperature bumps, sublimation lines and, therefore, planetary traps. They claimed that the in situ formation of close-in super-Earths would require canceling type I migration inside 1 au.

Our disk structure shows that it is possible for planets of a few Earth masses to a few tens of Earth masses to get trapped at various locations inside 2 au. Thus, assuming that these planets formed either in situ (yet beyond 0.2 au) or in the outer disk, they may survive if they do not accrete too much mass. After 1 million years of evolution, we found traps located at 0.3 (for planets up to $21 M_{\mathrm{\oplus}}$), 0.6 (up to $30 M_{\mathrm{\oplus}}$), 1.3(up to $49 M_{\mathrm{\oplus}}$), and 1.9 au (up to $40 M_{\mathrm{\oplus}}$), which translates into periodicity ratios of 2.83 between the inner most and the second trap, and 1.77 between the third one and the fourth one (the others ranging between 3 and 16). After 5 million years, the inner traps are located at 0.3, 0.5, 0.9, and 1.45 au (for planets up to $20, 25, 44,$ and $34 M_{\mathrm{\oplus}}$, respectively), which translates into periodicity ratios of 2.15 between the inner most and the second trap, 2.41 between the second and the third, and 2.04 between the third and the fourth.

Thus, periodicity ratios lower than 2 are consistent with the exoplanet observations that showed peaks at ratios of 2 and 5:3. Yet, we do not find any other correlation in periodicity ratios. Since our model only involves the viscous evolution of the disk and no real modeling of the individual planet dynamics, the mentioned correlation most certainly emerged by coincidence. However, the trap locations suggest that planets that would form further out in the disk and migrate towards these traps of periodicity ratios around 2, could be counted as in resonance.

Our simulations show that there is a possibility of retaining super-Earths (up to a few tens of $M_{\mathrm{\oplus}}$) that formed beyond 0.2 au in the simulations of \citet{ogihara15} by trapping them in our traps below 1 au. Though the second, third, and fourth traps can hold planets slightly more massive, the innermost trap can save planets up to $20 M_{\mathrm{\oplus}}$, which is consistent with the fact that the vast majority of detected super-Earths that are below $20 M_{\mathrm{\oplus}}$.

\section{Conclusions and perspectives}
\label{cclpersp}
\subsection{Conclusions}
Based on the viscous evolution of a MMSN, we have studied the possibility of trapping planets of various masses and preventing them from falling onto their host star. We have estimated the total torque exerted by the disk on a planet of a given mass at a given distance from the star. The mass of the planet mainly influences the saturation of the corotation torque, thereby affecting the positions of the planet traps and deserts across the disk.

For low mass ($M_{P} < 5 M_{\mathrm{\oplus}}$) or massive planets ($M_{P} \ge 100 M_{\mathrm{\oplus}}$), there appears to be only one possibility of long-term trapping, located at the location of the heat transition barrier around 10 au (varying between 8 and 20 au with time). Only intermediate-mass planets ($10 M_{\mathrm{\oplus}} \le M_{P} \le 50 M_{\mathrm{\oplus}}$) can get trapped at lower distance from the star: in these cases the traps usually coincide with the sublimation line of one of the major dust components (preferentially water ice, refractory organics, and silicates).

Trapped intermediate-mass planets are likely to remain trapped as they grow, until they reach the maximum mass of the outward migration zone where they are located. Then they might resume their inward migration, maybe get trapped to the next inner trap, and eventually fall onto the star. More massive trapped planets will also remain trapped in their respective traps slightly outer in the disk. If their mass and the local aspect ratio allow it, they might open a gap and then change from type I to type II migration. We estimate that massive planets are more likely to open gaps at certain trap locations, correlated with the edge of the shadowed regions: planets more massive than $170 M_{\mathrm{\oplus}}$ can open a gap around 15.8 au after 10 000 years of viscous evolution, or around 5.5 and 9 au after 5 million years.

We have identified several super-Earth traps in the disk, some of which are located around 5 and 7.5 au, though some others, associated with the sublimation lines, are located inside 2 au. These innermost traps would allow the survival of close-in super-Earths up to a few tens of Earth masses, such as the ones reported in exoplanet observations. They may also prevent the inward migration of super-Earths formed in situ below 1 au in the numerical simulations of \citet{hansen12}, \citet{hansen13} and \citet{ogihara15}. In addition, beyond the identification of resonance pairs in the planet periodicities, it would be very interesting to correlate these periodicity ratios with the trap locations.

Finally, observing young disks that possibly host planets and gaps (e.g. HL Tau), and coupling these observations with exoplanet detection and analysis would be very interesting to validate the giant planet-trapping at the heat transition barrier.

\subsection{Perspectives}
\label{persp}
By coupling the disk evolution with planetary growth, we may be able to model planet dynamics, and then estimate at what distance the planet is more likely to get trapped and possibly open a gap. Knowing how the planet mass grows will determine its dynamics and its chances of survival. Modeling the planetary growth consistently with the disk evolution will therefore be the object of a future study.

In addition, it becomes essential to take into account type II migration since gap openings are going have a very strong effect on the gas density distribution in the disk.

Moreover, the presence of multiple planets located at the same trap line (outer edge of the same outward migration zone, though with various masses) will alter the disk by inducing planet-planet resonant interactions, as well as disk-planet resonances, as related in \citet{cossou13}: planets can be trapped in chains of mean motion resonances. This requires a strong coupling between the hydrodynamical models of the disk evolution and the N-body models for planet interactions. It would be interesting to study the effect of this refinement on the counterreaction of the planet on the disk.

Although we used a detailed opacity model that accounts for the main dust components with a sublimation temperature above that of water ice, it might be interesting to add this opacity refinement for species with lower sublimation temperature to account for, for example, the presence of CO, CO$_{2}$, CH$_{4}$, and NH$_{3}$ ices.

Finally, recent studies by \citet{fouchet10} and \citet{gonzalez15}, using bi-fluid SPH simulations, tend to show that the dust-to-gas ratio cannot be considered as constant and uniform in presence of planets. Adapting the opacity model to reflect dust-to-gas ratio variations would be the object of a future study.

\begin{acknowledgements}

We thank Dominic Macdonald for valuable suggestions that improved the quality of the manuscript significantly. We also thank the referee for detailed and constructive comments that improved the quality of the paper. This work was supported by IDEX Sorbonne Paris Cit\'e and the Conseil Scientifique de l'Observatoire de Paris. We acknowledge the financial support from the UnivEarthS Labex program of Sorbonne Paris Cit\'e (ANR-10-LABX-0023 and ANR-11-IDEX-0005-02) and from the Institut Universitaire de France.  
\end{acknowledgements}


\bibliography{bibliography}
\end{document}